\documentclass[twocolumn,floatfix,prl,aps,showpacs,superscriptaddress]{revtex4-2}
\usepackage{graphicx,amsmath,amssymb,color,nicefrac,multirow, makecell, ragged2e,float,multirow,arydshln}
\usepackage[unicode=true,colorlinks=true,linkcolor=blue,citecolor=blue,urlcolor=blue]{hyperref}
\usepackage[titletoc,title]{appendix}

\newcommand{\be}{\begin{equation}}
\newcommand{\ee}{\end{equation}}

\newcommand{\ba}{\begin{eqnarray}}
\newcommand{\ea}{\end{eqnarray}}

\renewcommand{\vec}[1]{{\bm{#1}}}

\def\beq{\begin{eqnarray}}
\def\eeq{\end{eqnarray}}

\makeatletter
\newcommand*{\rom}[1]{\expandafter\@slowromancap\romannumeral #1@}
\makeatother

\newcommand{\non}{\nonumber\\}

\newcommand{\pa}{\partial}
\usepackage{bm,physics,enumerate,booktabs,ulem}

\newcommand{\tx}[1]{{\text{#1}}}

\newcommand{\lB}{l_B}
\newcommand{\omegac}{\omega_c}
\newcommand{\s}{\mathfrak{s}_{\theta}}

\usepackage{orcidlink}

\begin{document}
\title{
Repulsive-Interaction-Driven Topological Superconductivity 
in a Landau Level Coupled to an $s$-Wave Superconductor
}

\author{Koji Kudo}
\affiliation{Department of Physics, Kyushu University, Fukuoka 819-0395, Japan}

\author{Ryota Nakai}
\affiliation{RIKEN Center for Quantum Computing (RQC), Wako, Saitama, 351-0198, Japan}

\author{Hiroki Isobe}
\affiliation{Department of Physics, Kyushu University, Fukuoka 819-0395, Japan}

\author{J. K. Jain\orcidlink{0000-0003-0082-5881}}
\affiliation{Department of Physics, 104 Davey Lab, The Pennsylvania State University, University Park, Pennsylvania 16802, USA}

\author{Kentaro Nomura}
\affiliation{Department of Physics, Kyushu University, Fukuoka 819-0395, Japan}
\affiliation{Quantum and Spacetime Research Institute, Kyushu University, Fukuoka 819-0395, Japan}

\begin{abstract}
A two-dimensional topologically nontrivial state of noninteracting 
electrons, such as the surface state of a three-dimensional topological
insulator, is predicted to realize a topological superconductor when 
proximity-coupled to an ordinary $s$-wave superconductor.  
In contrast, noninteracting electrons partially occupying a Landau 
level, with Rashba spin-orbit coupling that lifts the spin degeneracy, fail to 
develop topological superconductivity under similar proximity coupling in the 
presence of the conventional Abrikosov vortex lattice.
We demonstrate, through exact diagonalization, that introducing in this model 
a repulsive interaction between electrons induces topological superconductivity
at half-filled Landau level for a range of parameters. This appears rather 
surprising 
because a repulsive interaction is expected to inhibit, not promote, pairing, 
but suggests an appealing principle for realizing topological 
superconductivity: proximity-coupling a composite Fermi liquid to an ordinary 
$s$-wave superconductor.
\end{abstract}

\maketitle

\paragraph{Introduction.}---
Realizations of topological superconductors would be of great interest as they are predicted to support Majorana modes, which are a prominent example of non-Abelian anyons. The Majorana particles were first envisioned, in the modern context, to appear either at the edges or inside the Abrikosov vortices of even-denominator fractional quantum Hall states, which are ``topological superconductors'' of composite fermions (CFs)~\cite{Jain89,Moore91,Greiter91,Wen91,Read96,Read00,Morf98,Park98,Jain07,Sharma21}. Since the first even-denominator fraction to be observed, namely $\nu=5/2$~\cite{Willett87}, many additional even-denominator fractional quantum Hall states ($\nu=1/2$, 1/4, 1/6, 1/8, 3/4, 3/8, 3/10, $2+3/8$) have been observed in semiconductor quantum wells~\cite{Suen92, Suen92b,Pan08,Shabani09a, Shabani09b, Kumar10,Bellani10,Drichko19,Wang23, Wang22a, Wang23b,Wang25a} and in bilayer~\cite{Zibrov16, Li17, Huang21, Assouline23, Kumar24} and trilayer graphene~\cite{Chen24,Chanda25}. These are understood in terms of $p$- or $f$-wave superconductivity (SC) of CFs~\cite{Mukherjee12, Mukherjee14c, Hutasoit16,Mukherjee15b,Balram18,Kim19,Faugno19, Zhu20a,Balram21,Balram21b,Sharma22,Sharma23,Zhao23}. We note that at the half-filled lowest Landau level (LL) in narrow quantum wells, CFs form a composite-Fermi liquid (CFL) state~\cite{Halperin93, Halperin20, Halperin20b, Shayegan20}, but the CFL is unstable to a pairing of CFs when the strength of the electron-electron interaction reduced by making the quantum well wider~\cite{Sharma23} or by enhancing LL mixing~\cite{Zhao23}. 

While nature has been generous with topological SC (TSC) of CFs, it has not 
given us a natural candidate for 
an intrinsic TSC of {\it electrons}.  Proposals have been made that such SC can
be engineered in a heterostructure which couples a topological 2D electron 
system with an ordinary $s$-wave SC
~\cite{Alicea12,Fu08,Akhmerov09,Sau10,Lutchyn10,Oreg10,Alicea10,Choy11,Nadj-Perge13,Braunecker13,Klinovaja13,Vazifeh13,Pientka13,Nakosai13,Nadj-Perge14,Qi10,Bjorn16,Jeon19,Mishmash19,Chaudhary20,Schirmer22,Schirmer24,Kudo24a,Nakai25,Antonenko25}.
Given that a LL is a topological band with a non-zero Chern number, one might 
expect that coupling it to an $s$-wave superconductor (typically a 
type-II to withstand the strong magnetic field) along with Rashba spin-orbit 
coupling, which lifts the spin degeneracy, might produce TSC. 
However, that turns out not to be the case in the simplest 
model~\cite{Mishmash19,Chaudhary20}.
To obtain TSC in such a system, one needs to introduce additional
ingredients, for example
an external periodic potential~\cite{Mishmash19,Schirmer24}, 
or an unconventional Abrikosov lattice~\cite{Chaudhary20,Schirmer22}, or 
disorder~\cite{Kudo24a}.

\begin{figure}[b]
\includegraphics[width=\columnwidth]{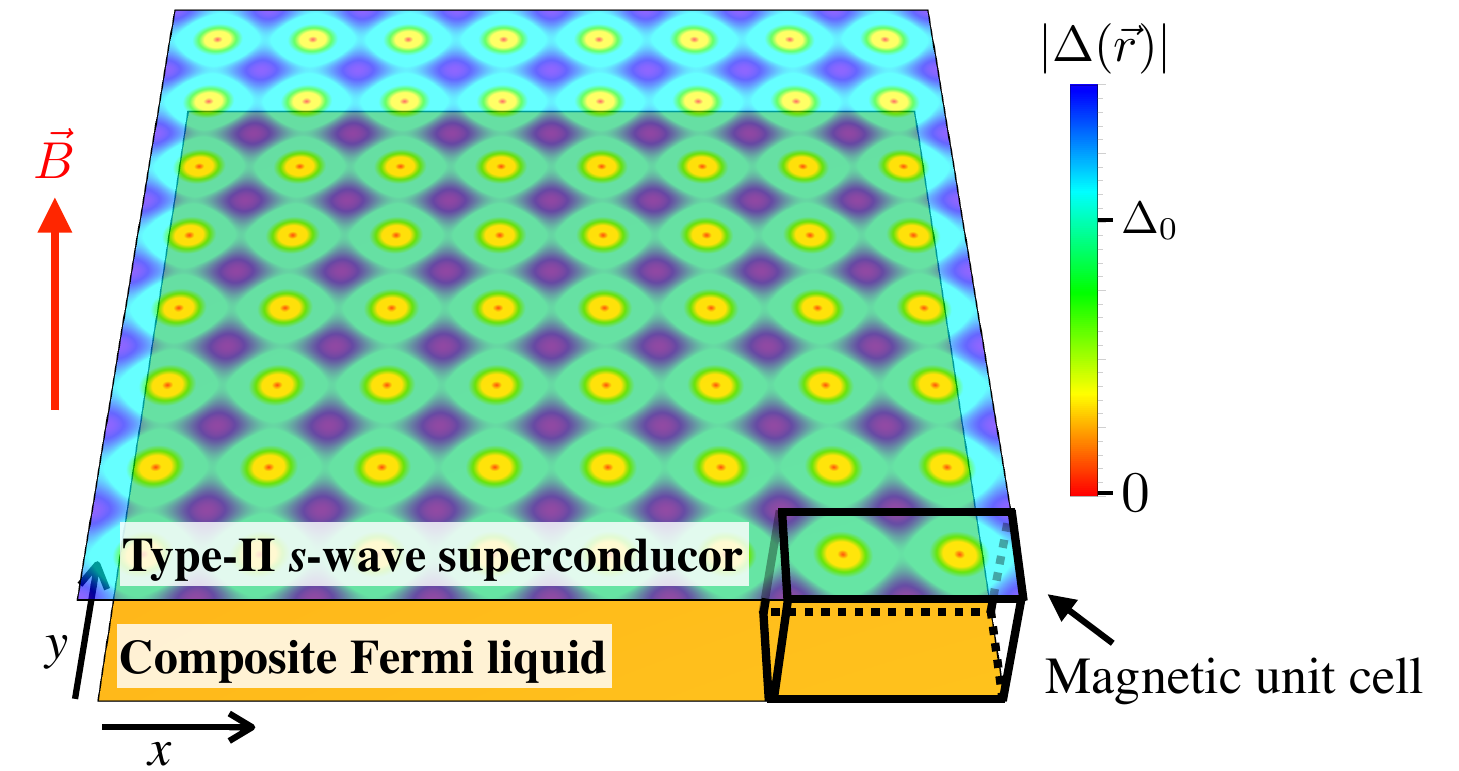}
 \caption{
 Schematic illustration of the hybrid system considered here. 
 An interacting electron system that forms a composite Fermi liquid (bottom 
 layer) is proximity-coupled to the type-II $s$-wave superconductor with an 
 Abrikosov vortex lattice (top layer).
 The cuboid
 outlines the magnetic unit cell (MUC) containing two vortices. 
 The system 
 shown corresponds to the largest
 size used in our exact diagonalization, consisting of 
 $n_x\times n_y=4\times8$ MUCs with total magnetic flux $N_\phi=32$.
 }
 \label{fig:hybrid}
\end{figure}

In this Letter, we show a conceptually simple
pathway to produce TSC. We demonstrate that repulsive electron-electron (e-e) interactions induce TSC at half-filling in a 
Rashba-coupled LL in proximity to a type-II  $s$-wave superconductor.
This is counter-intuitive, as one would expect a repulsive e-e interaction to 
be antithetical to SC. We refer to it as the repulsive-interaction-driven TSC 
(RID-TSC). 
As the strength of the repulsive e-e interaction is increased, the 
system eventually transitions into a CFL. This observation 
invites us to view this problem as that of the CFL proximity-coupled to an 
$s$-wave superconductor (see Fig.\ref{fig:hybrid}), providing a new 
design principle for realizing TSC.
In this study, we perform an extensive exact diagonalization by placing 
the system on a torus to properly deal with the e-e interaction.
It is likely that in typical systems the repulsive e-e interaction is stronger than the strength of the pairing potential. The above result suggests that producing TSC in this hybrid
would require a weakening or screening of the e-e interaction, say by nearby screening layers~\cite{Nakamura20}. 

\paragraph{Model.}--- 
We consider a two-dimensional system of interacting spinful electrons 
with Rashba spin-orbit coupling.
Periodic boundary conditions are imposed
in both $x$ and $y$ directions.
The electrons are subject to a perpendicular magnetic field
and proximitized by an type-II $s$-wave superconductor (Fig.~\ref{fig:hybrid}).
The total Hamiltonian reads
$H_\tx{hybrid}=H_0+H_\tx{int}+H_\Delta-\mu N$, where
\begin{align}
 &H_0
 =\int d^2r
 c^\dagger(\vec{r})
 \left[
 \frac{\vec{\pi}^2}{2m_e}
 -\alpha_R\left(\vec{\sigma}\times\vec{\pi}\right)_z\right]
 c(\vec{r}),\non
 &H_\tx{int}
 =\frac{1}{2}\sum_{\sigma\sigma'}\int d^2rd^2r'\,
 c_{\sigma}^\dagger(\vec{r})c_{\sigma'}^\dagger(\vec{r}')V(|\vec{r}-\vec{r}'|)
 c_{\sigma'}(\vec{r}')c_{\sigma}(\vec{r}),\non
 &H_\Delta
 =\int d^2r
 \left[
 c^\dagger_\uparrow(\vec{r})\Delta(\vec{r})c^\dagger_\downarrow(\vec{r})
 +\tx{h.c.}\right].
 \label{eq:originalH}
\end{align}
Here,
$c^\dagger(\vec{r})
=(c^\dagger_\uparrow(\vec{r}),c^\dagger_\downarrow(\vec{r}))$ is
the electron field operator, $\vec{\pi}$ is the canonical momentum, $m_e$ is 
the electron mass, $\vec{\sigma}$ is the Pauli matrices, and $\alpha_R$ is the 
Rashba coupling strength. 
We consider a screened Coulomb potential~\cite{screen}: $V(r)=V_Ce^{-r/\lB}/(r/\lB)$, where $\lB=\sqrt{\hbar c/|e|B}$ is the magnetic length and $V_C$
is the interaction strength. $\Delta(\vec{r})$ is the 
proximity-induced 
pair potential, $\mu$ is the chemical potential, and $N$ is the number 
operator. 
We work slightly below the upper critical field $H_{c2}$ of the superconductor,
where
$\Delta(\vec{r})$ forms an Abrikosov vortex lattice. Here, we assume a square 
vortex lattice as
$\Delta(\vec{r})=\left(\Delta_0/\sqrt{2}\right)
\sum_{j}\varphi^\tx{LLL}_{j}(\vec{r})$, where $\Delta_0$ is the pairing 
strength and $\varphi^\tx{LLL}_{j}(\vec{r})$ is the lowest LL wavefunction of 
charge-$2e$ Cooper pairs with momentum index $j$~\cite{Abrikosov}.

Due to the large penetration depth close to $H_{c2}$, 
the electrons 
experience an approximately 
uniform magnetic field, leading to Landau quantization. 
The LLs are spin-split by Rashba coupling and the spectrum becomes
$\epsilon_{n,\tau}
=\hbar\omegac\left(n+\tau\sqrt{1/4+g_R^2n}\right)$ with 
$n=0,1,\ldots$, where $\tau=\pm1$ labels the two Rashba-split branches,
$\omegac=|e|B/m_ec$ is the 
cyclotron frequency, and $g_R=\sqrt{2}\alpha_R/\omegac\lB$ is the dimensionless
Rashba coupling strength.
We adopt a rectangular magnetic unit cell (MUC)~\cite{Mishmash19} as shown in 
Fig.~\ref{fig:hybrid}. A system with $n_x\times n_y$ MUCs then contains 
$N_\phi=n_xn_y$ magnetic flux quanta~\cite{flux}.
Each LL accommodates
$N_\phi$ single-particle states labeled by the Bloch momentum 
$\vec{k}=(k_x,k_y)=2\pi(j_x/2n_x,j_y/n_y)$ with
$j_\alpha=0,1,\ldots,n_\alpha-1$, where the intervortex separation is set to 
unity (see Sec.~\ref{sec:2d} in Supplemental Material (SM)~\cite{SM} for more
details on vortex lattices, Rashba-coupled LLs, and the Bloch basis).

We project the Hamiltonian onto the lowest Rashba-coupled LL with energy
$\epsilon_{1,-1}$, assuming that $\mu$ is tuned near this LL and both
the interaction energy and the pair potential are weak compared to the LL 
spacing. This yields an effective Hamiltonian
$\tilde{H}_\tx{hybrid}=\tilde{H}_\tx{int}+\tilde{H}_\Delta-\mu N$, where
\begin{align}
 \tilde{H}_\tx{int}
 &=\sum_{\vec{k}_1\vec{k}_2\vec{k}'_1\vec{k}'_2}
 V_{\vec{k}_1\vec{k}_2\vec{k}'_1\vec{k}'_2}
 c_{\vec{k}_1}^\dagger c_{\vec{k}_2}^\dagger c_{\vec{k}'_2}c_{\vec{k}'_1},
 \label{eq:Hint}\\
 \tilde{H}_\Delta
 &=\sum_{\vec{k}}
 \Delta_{\vec{k}}
 c_{\vec{k}}^\dagger c_{-\vec{k}}^\dagger+\tx{h.c.}
 \label{eq:HDelta}
\end{align}
Here $ c_{\vec{k}}^\dagger$ is the creation operator in the projected LL. 
The explicit forms of the matrix elements 
$V_{\vec{k}_1\vec{k}_2\vec{k}'_1\vec{k}'_2}$ and 
$\Delta_{\vec{k}}$ are given in Sec.~\ref{sec:matrix} in 
SM~\cite{SM}. 
For convenience, we simplify the treatment of the Rashba coupling strength
$g_R$ as described in the footnote~\cite{gR}, so that it does not 
explicitly 
appear hereafter. The system is therefore governed by three parameters: 
the interaction strength $V_C$, the pairing strength $\Delta_0$,
and the chemical potential $\mu$.

In this study, 
we perform exact diagonalization of the projected Hamiltonian 
$\tilde{H}_\tx{hybrid}$ in the
full Fock space, spanning all particle-number sectors from $N=0$ to $N=N_\phi$.
This method treats repulsive interactions and superconducting pairing 
simultaneously, providing a powerful tool for studying hybrid systems of this 
kind.
The Hamiltonian $\tilde{H}_\tx{hybrid}$ conserves the total momentum $\vec{K}$
and the fermion parity $P\equiv(-1)^N$. Within each $(\vec{K},P)$ subspace, 
we employ the Lanczos algorithm to obtain the low-energy 
spectrum~\cite{Lanczos}.
Unless otherwise stated, we focus on the half-filling defined by 
$\nu\equiv\langle N\rangle/N_\phi=1/2$ by tuning $\mu$, where 
$\langle\cdot\rangle$ denotes the ground-state (in the Fock space) expectation value.

\begin{figure}[t]
\includegraphics[width=\columnwidth]{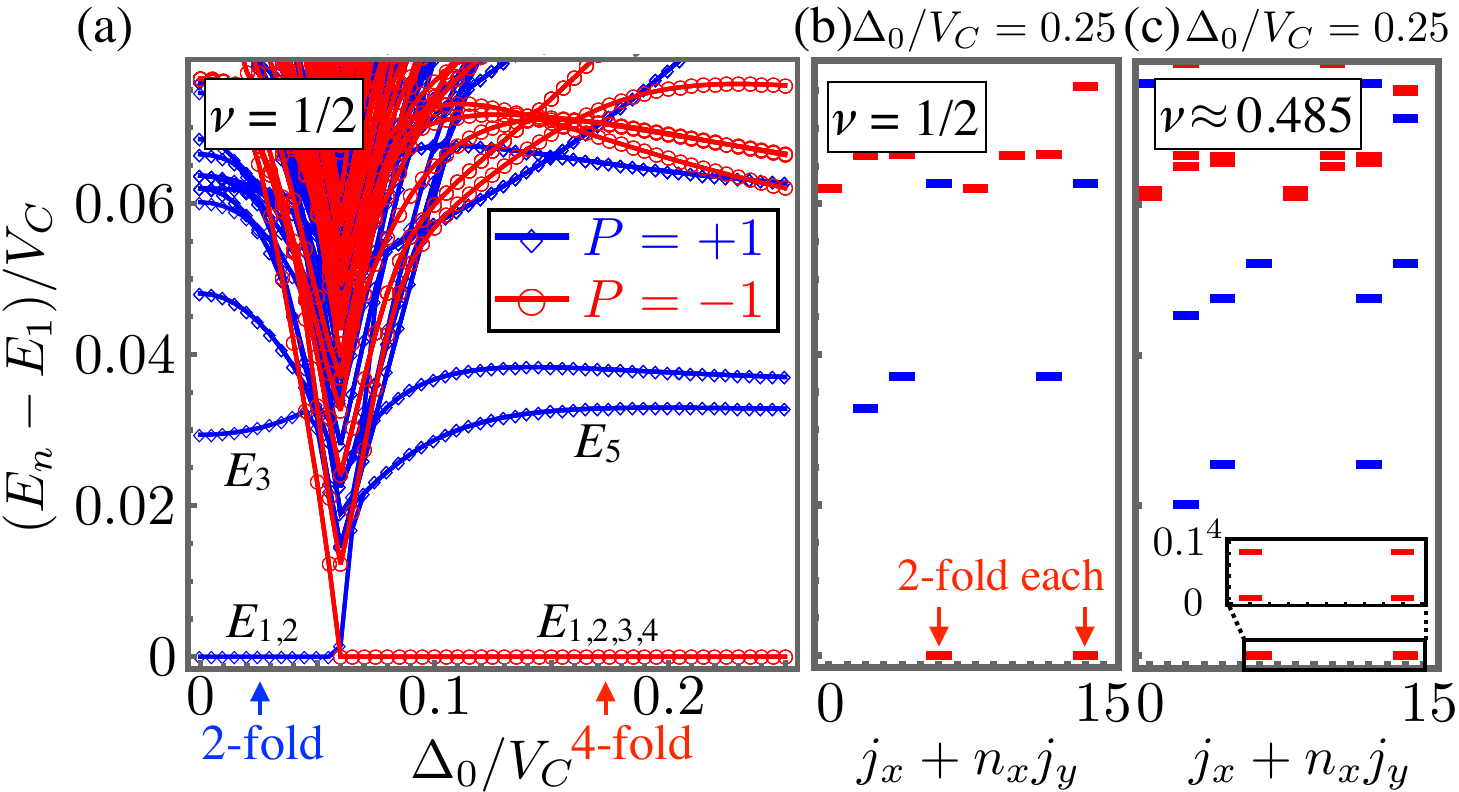}
 \caption{
 (a) Low-energy spectrum
 $E_n-E_1$ at $\nu=1/2$ as a function of the pairing 
 strength $\Delta_0$, where $E_n$ is the $n$th lowest energy. Energies are 
 measured in units of the interaction strength $V_C$. 
 Colors indicate the fermion parity $P$. The system size is 
 $N_\phi=n_x\times n_y=4\times4$.
 A transition occurs
 near $\Delta_0/V_C\approx0.06$.
 (b) Spectrum at $\Delta_0/V_C=0.25$, plotted versus total momentum index
 $j_x+n_xj_y$, where $j_\alpha=0,1,\ldots,n_\alpha-1$. The ground state 
 consists of two states at $\vec{K}=(\pi/2,\pi/2)$ and two at 
 $\vec{K}=(\pi/2,3\pi/2)$.
 (c) Same as (b), but at $\nu\approx0.485$. The inset shows a lifting of the 
 twofold degeneracy within each momentum sector.
 }
 \label{fig:CFStoSC}
\end{figure}

\paragraph{Induced-superconductivity in the Rashba-coupled LL.}--- 
We begin by showing that
proximity-induced pairing drives a quantum phase transition
from a CFL into a superconductor. 
Figure~\ref{fig:CFStoSC}(a) presents the low-energy spectrum $E_n-E_1$
($E_n$ is the $n$th lowest energy) 
as a function of $\Delta_0$. 
The ground state at $\Delta_0/V_C=0$ corresponds to a CFL with a finite-size gap~\cite{RashbaCF} characterized by twofold degeneracy and even fermion parity 
$P=+1$.
As $\Delta_0$ increases, a level crossing occurs at finite
$\Delta_0$, indicating a transition to a new ground state characterized by:
\begin{enumerate}[(i)]
 \setlength{\parskip}{0cm}
 \setlength{\itemsep}{0cm}
 \item fourfold degeneracy, \label{item:degeneracy}
 \item total momentum $\vec{K}=(\pi/2,\pi/2)$ and $(\pi/2,3\pi/2)$ 
       (two states at each),
       \label{item:momentum}
 \item odd fermion parity ($P=-1$). \label{item:parity}
\end{enumerate}
Features~(\ref{item:degeneracy}) and (\ref{item:momentum}) 
are confirmed in the spectrum at $\Delta_0/V_C=0.25$ in 
Fig.~\ref{fig:CFStoSC}(b).
The $\vec{K}=(\pi/2,\pi/2)$ and $(\pi/2,3\pi/2)$ states
are degenerate 
and related by the symmetry operation $\Pi R_z(\pi)$, with
$\Pi$ the inversion and $R_z(\pi)$ the $\pi$ spin-rotation about the $z$ axis.
To examine the origin of the twofold degeneracy at each $\vec{K}$ 
sector,
we compute the spectrum slightly away from $\nu=1/2$ 
[Fig.~\ref{fig:CFStoSC}(c)]. The twofold degeneracy is lifted, indicating that 
it is protected by electron-hole symmetry at $\nu=1/2$.
Feature~(\ref{item:parity}) will be discussed below in relation to TSC.

\begin{figure}[t]
\includegraphics[width=\columnwidth]{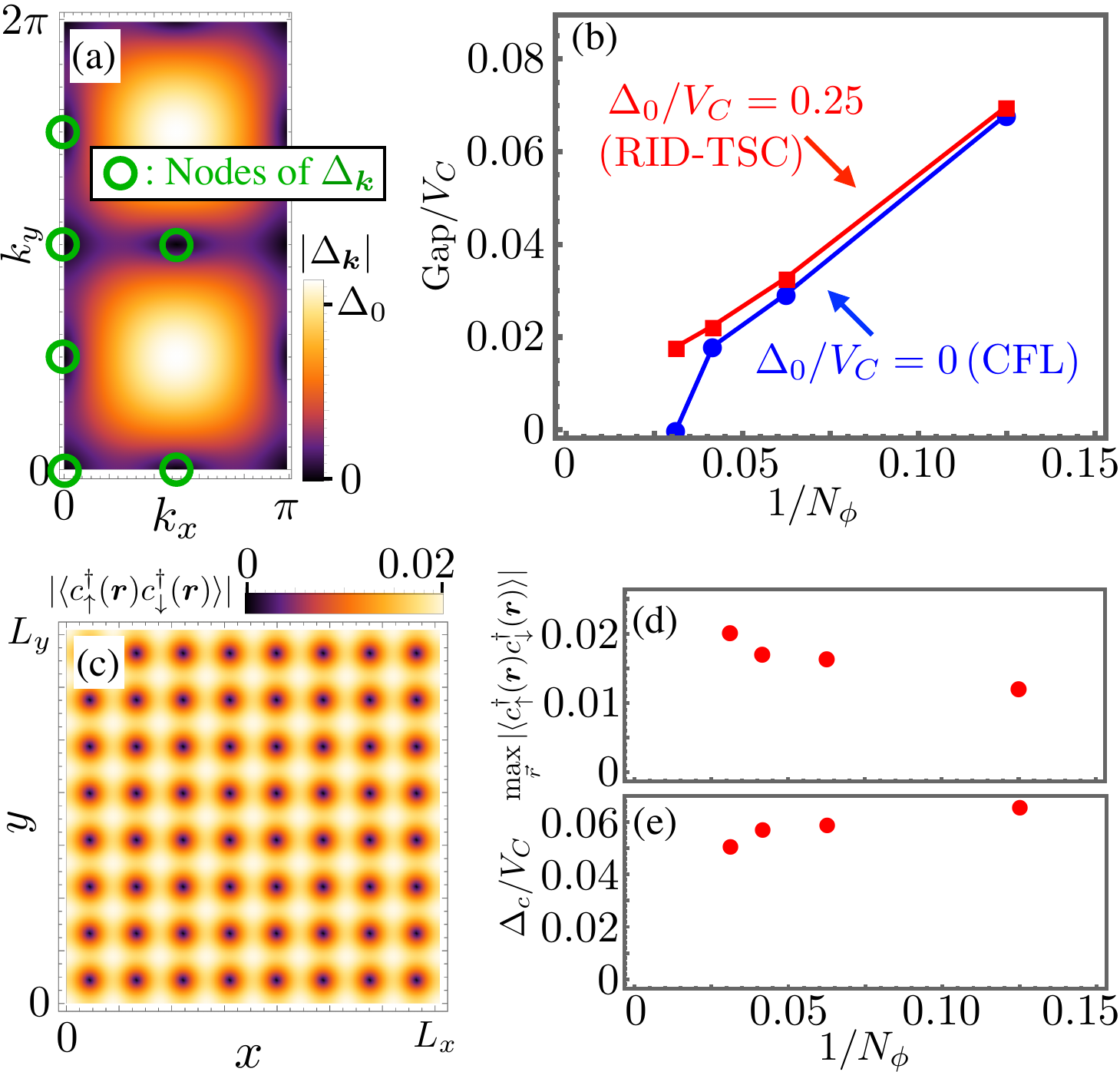}
 \caption{
 (a) Modulus of $\Delta_\vec{k}$.
 The circles indicate nodes.
 (b) Finite-size scaling of the energy gaps for the fourfold degenerate
 RID-TSC
 state at $\Delta_0/V_C=0.25$ and for the CFL state at 
 $\Delta_0/V_C=0$.
 (c) Spatial profile of 
 $|\langle c^\dagger_\uparrow(\vec{r})c^\dagger_\downarrow(\vec{r})\rangle|$ at
 $\Delta_0/V_C=0.25$ and $n_x\times n_y=4\times8$. $L_{x(y)}$ denotes the
 system length. 
 (d,e) Finite size scaling of (d)
 $\max_{\vec{r}}|\langle c^\dagger_\uparrow(\vec{r})c^\dagger_\downarrow(\vec{r})\rangle|$ 
 and (e) the critical pairing strength $\Delta_c/V_C$, defined as the
 $\Delta_0/V_C$ where the phase transition to the RID-TSC phase occurs.
 }
 \label{fig:Finite}
\end{figure}

Within system sizes accessible in our study, the ground state features listed 
in (\ref{item:degeneracy})-(\ref{item:parity}) appear at $\nu=1/2$.
Here, we restrict system sizes to be
$n_x\times n_y=2s\times 4t\,(s,t\in \mathbb{N})$,
where the discrete momenta include all point nodes of $\Delta_{\vec{k}}$ 
defined in Eq.~\eqref{eq:HDelta}, which
govern the low-energy physics [see Fig.~\ref{fig:Finite}(a) plotting the 
modulus of $\Delta_\vec{k}$].
In particular, we studied $(n_x,n_y)=(2,4),(4,4),(6,4),$ and 
$(4,8)$~\cite{NextSize}.
The energy spectra as in Fig.~\ref{fig:CFStoSC}(a) for various $(n_x,n_y)$, 
including those other than the above list, are provided in Sec.~\ref{sec:nxny} 
in SM~\cite{SM}.

We call the state with
Features~(\ref{item:degeneracy})-(\ref{item:parity})
repulsive-interaction-driven topological
superconductivity (RID-TSC) based on its nature, which we will demonstrate 
below.

\paragraph{Finite-size scaling analysis} ---
We first perform finite-size 
scaling using the system sizes described above. 
In Fig.~\ref{fig:Finite}(b), we plot the energy gaps of the RID-TSC state
and the CFL state as functions of $1/N_\phi$. Unlike the CFL, whose gap 
collapses 
rapidly with the system size, the RID-TSC gap
decreases more slowly, with extrapolation suggesting a finite value in the 
thermodynamic limit.
A more quantitative assessment will require study of larger systems, 
which may be accomplished, for example, by the density matrix renormalization 
group. Such a study is beyond the scope of the present work, however.

To address SC, we compute the $s$-wave channel of
the pair amplitude
$\langle c^\dagger_\uparrow(\vec{r})c^\dagger_\downarrow(\vec{r})\rangle$,
which shows the Abrikosov vortex [Fig.~\ref{fig:Finite}(c)].
Figure~\ref{fig:Finite}(d) plots 
the maximum modulus
$\max_{\vec{r}}|\langle c^\dagger_\uparrow(\vec{r})c^\dagger_\downarrow(\vec{r})\rangle|$
versus $1/N_\phi$,
indicating that a finite pair amplitude is introduced.
The other components of the induced pair amplitude will be discussed later.

We also plot in Fig.~\ref{fig:Finite}(e) 
the finite-size scaling of the critical pair potential $\Delta_c$
where the phase transition to the RID-TSC phase occurs.
The present calculations suggest finite $\Delta_c$; should this remain
the case in the thermodynamic limit, the RID-TSC would be distinct from a 
paired state of CFs;
if the proximity effect were to induce pairing of CFs, such a state would 
likely emerge
at infinitesimal $\Delta_0$, reflecting the instability of the CFL. 
This interpretation is also supported by Fig.~\ref{fig:CFStoSC}(a),
where near $\Delta_c$ many low-energy states simultaneously move down in 
energy to drastically reorganize the spectrum. Such behavior implies a 
transition between qualitatively distinct phases, consistent with the RID-TSC 
not being a paired state of CFs.
Indeed, our
phase exhibits fourfold ground state degeneracy, in contrast
to the sixfold degeneracy~\cite{Read00,Oshikawa07} of the Moore-Read state.
Nevertheless, the ground state degeneracy may suggest
topological order, though its identification requires additional work, such as
computing the topological entanglement entropy.

\begin{figure}[t]
\includegraphics[width=\columnwidth]{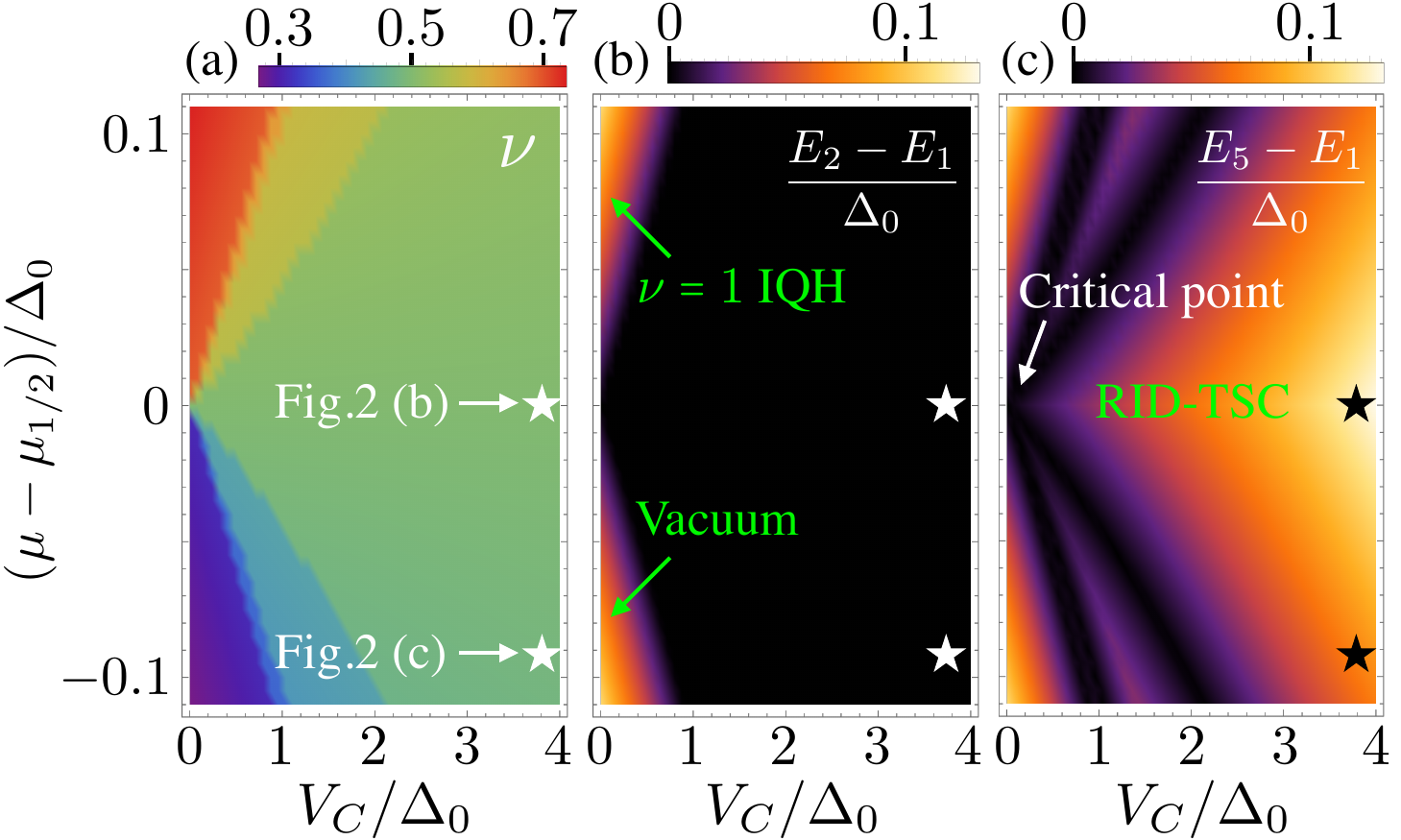}
 \caption{
 (a) Filling factor $\nu$ and (b)(c) energy differences
 $(E_n-E_1)/\Delta_0$ for $n=2$ and $n=5$ respectively, as functions of the interaction 
 strength $V_C/\Delta_0$ and
 $(\mu-\mu_{1/2})/\Delta_0$, where $\mu_{1/2}$ is the chemical potential 
 yielding $\nu=1/2$. Stars mark the parameters corresponding to 
 Figs.~\ref{fig:CFStoSC}(b) and \ref{fig:CFStoSC}(c). 
 The gapped regions in (b) corresponds to 
 vacuum and $\nu=1$ IQH phases, respectively.
 The RID-TSC phase in (c) emerges from the critical point 
 $V_C/\Delta_0=(\mu-\mu_{1/2})/\Delta_0=0$ and expands as
 $V_C/\Delta_0$ increases. The system size is 
 $n_x\times n_y=4\times4$. The composite Fermi liquid appears at 
 $V_C/\Delta_c\approx17$, beyond the plotted range.
 }
 \label{fig:Many-body}
\end{figure}

\paragraph{Repulsive interaction and superconductivity} ---
Next we demonstrate the interaction-driven nature of the RID-TSC phase. 
Figure~\ref{fig:Many-body} shows the filling factor $\nu$ and 
the energy differences $(E_n-E_1)/\Delta_0$ for $n=2$ and $n=5$,
as 
functions of $V_C/\Delta_0$ and $(\mu-\mu_{1/2})/\Delta_0$, where $\mu_{1/2}$ 
is the chemical potential yielding $\nu=1/2$. 
In Fig.~\ref{fig:Many-body}(b), two gapped regions
appear at $V_C/\Delta_0\approx0$: one for $\mu-\mu_{1/2}<0$, adiabatically
connected to the vacuum state (i.e. $\nu=0$), and another for 
$\mu-\mu_{1/2}>0$, connected to the $\nu=1$ IQH state. 
This indicates that for noninteracting electrons only these two phases occur,
with no possibility of TSC, as shown in Ref.~\onlinecite{Mishmash19}.

The fourfold degenerate state in the RID-TSC phase appears as a broad gapped 
region in Fig.~\ref{fig:Many-body}(c). 
Strikingly, it emerges from the 
{\it critical point at $V_C/\Delta_0=0$} between
the vacuum and the $\nu=1$ IQH phases, and significantly expands as 
$V_C/\Delta_0$ increases. We note that the RID-TSC phase emerges once the 
repulsive interaction is made finite, yet disappears in noninteracting limit.
This underscores its repulsive-interaction-driven 
origin. 
If $V_C/\Delta_0$ exceeds the plotted range in 
Fig.~\ref{fig:Many-body}(c), the system
eventually undergoes a transition into the CFL as demonstrated in 
Fig.~\ref{fig:CFStoSC}(a).
(We here regard the four lowest-energy states as the ground states of the 
RID-TSC phase.
To be precise, the degeneracy is lifted when $\nu$ deviates from 1/2 but the deviation is much smaller than $E_5-E_1$, see Sec.~\ref{sec:gap3-1} in 
SM~\cite{SM}.)

\paragraph{Topological superconductivity.}---
The RID-TSC phase exhibits odd fermion parity, $P=-1$ 
[Feature~\eqref{item:parity}]. We now use this property to diagnose its 
topological character.

First, consider a generic noninteracting Bogoliubov-de Gennes (BdG) 
Hamiltonian for 
two-dimensional spinless fermions:
$H_\tx{BdG}=\sum_{\vec{k}}\left(\epsilon_\vec{k}-\mu\right)c_\vec{k}^\dagger c_\vec{k}+\sum_\vec{k}\Delta_\vec{k}c_\vec{k}^\dagger c_{-\vec{k}}^\dagger+\tx{h.c.}$, where
$\epsilon_\vec{k}$ is the single-particle dispersion (assumed, 
$\epsilon_{-\vec{k}}=\epsilon_\vec{k}$ and 
$\Delta_{-\vec{k}}=-\Delta_\vec{k}$).
Introducing quasiparticle operators
$\alpha_\vec{k}^\dagger=u_\vec{k}c^\dagger_\vec{k}+v_\vec{k}c_{-\vec{k}}$,
that diagonalizes $H_\tx{BdG}$,
one obtains the ground state~\cite{Read00} 
\begin{align}
 \ket{\Omega}
 =\prod_{\vec{k}\neq\vec{k}^*}{}'
 \left(u_\vec{k}^*-v_\vec{k}^*c_\vec{k}^\dagger c_{-\vec{k}}^\dagger\right)
 \prod_{\epsilon_{\vec{k}^*}-\mu<0}c^\dagger_{\vec{k}^*}\ket{0},
 \label{eq:Omega}
\end{align}
where $\prod{}'_{\vec{k}\neq\vec{k}^*}$ runs over distinct
$(\vec{k},-\vec{k})$ pairs, excluding $\vec{k}^*$, and $\vec{k}^*$ denote
nodes of $\Delta_{\vec{k}}$. (Here we assume that $\vec{k}$ is continuous; for
discrete $\vec{k}$, we take system sizes whose allowed momenta include 
$\vec{k}^*$).
Equation~\eqref{eq:Omega} implies that the fermion parity $P$ of $\ket{\Omega}$
is determined solely by the 
occupations at $\vec{k}^*$. From this property, one can show that $P$ coincides
with the parity of the BdG Chern number $\mathcal{N}$ of $\ket{\Omega}$:
(see Sec.~\ref{sec:FP} in SM~\cite{SM} for more details)
\begin{align}
 P=(-1)^{\mathcal{N}}.
 \label{eq:PNformula}
\end{align}
Equation~\eqref{eq:PNformula} implies that any gapped BdG
state with $P=-1$ necessarily exhibits TSC with an odd $\mathcal{N}$. This can 
be understood by a boundary argument: a $P=-1$ SC state and a topologically 
trivial one belong to distinct fermion-parity sectors, and thus 
the junction between them must exhibit gap closing at 
the boundary. 
This results in gapless Majorana edge modes as dictated by the odd
$\mathcal{N}$ associated with $P=-1$.

Our interacting Hamiltonian $\tilde{H}_\tx{hybrid}$, obtained from
$H_\tx{BdG}$ by discarding the kinetic term and adding interactions, still 
preserves $P$.
Therefore, we expect
the above boundary argument to hold, 
indicating that the RID-TSC phase
characterized by $P=-1$ exhibits 
topological superconductivity with gapless Majorana edge modes.

Equation~\eqref{eq:Omega}
further shows that the RID-TSC phase
cannot be adiabatically connected to any mean-field state $\ket{\Omega}$,
reflecting its interaction-driven origin. 
The total momentum of $\ket{\Omega}$ is given by
$\vec{K}=\sum_{\epsilon_{\vec{k}^*}-\mu<0}\vec{k}^*$.
Moreover,
inversion symmetry in $H_\tx{BdG}$ constrains 
it to inversion-invariant momenta 
(IIMs), for which $\vec{K}=-\vec{K}$ modulo the Brillouin zone. The IIMs 
in our hybrid system are
$(0,0)$, $(\pi/2,0)$, $(0,\pi)$, and $(\pi/2,\pi)$. In contrast,
the RID-TSC phase has $\vec{K}=(\pi/2,\pi/2)$ and $(\pi/2,3\pi/2)$ [Feature
(\ref{item:momentum})], which rules out any adiabatic deformation to any
mean-field state $\ket{\Omega}$.
In particular, one of the additional nodes at $(0,\pi/2)$ 
and $(0,3\pi/2)$ [see Fig.~\ref{fig:Finite}(a)], which are not included in the 
IIMs, must be occupied to account for $\vec{K}=(\pi/2,\pi/2)$ and 
$(\pi/2,3\pi/2)$.

\begin{figure}[t]
\includegraphics[width=\columnwidth]{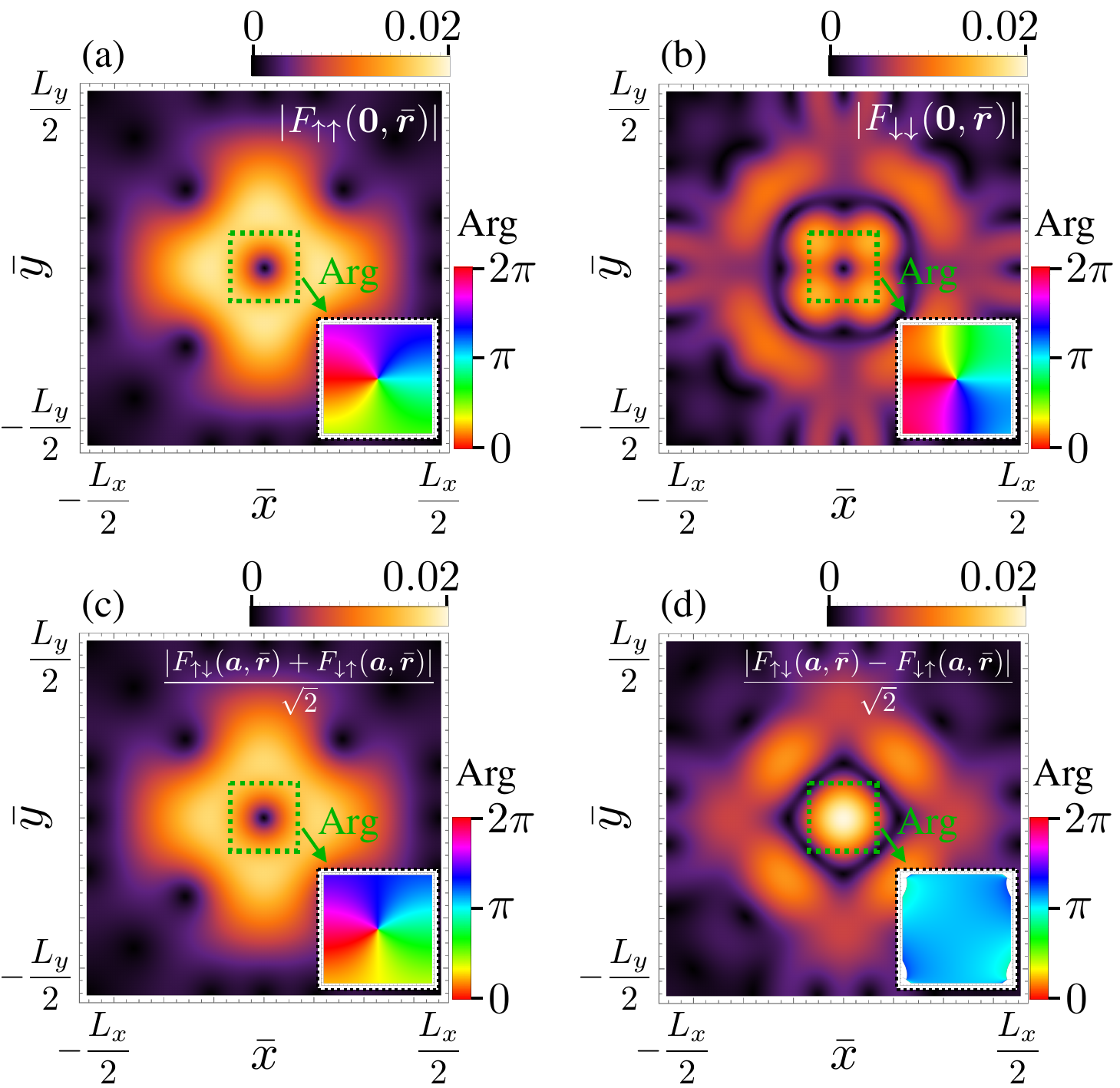}
 \caption{
 Modulus of the pair amplitude
 $F_{\sigma_1\sigma_2}(\vec{R},\bar{\vec{r}})$. Panels show
 (a) $s_z=1$, 
 (b) $s_z=-1$, (c)$s_z=0$ for $p$-wave and (d)$s_z=0$ for $s$-wave pairing,
 plotted as a function of the relative coordinate 
 $\bar{\vec{r}}=(\bar{x},\bar{y})$. The 
 center-of-mass coordinate is set as (a)(b)$\vec{R}=\vec{0}$ and 
 (c)(d) $\vec{R}=\vec{a}\equiv\lB/2(1,1)$. The insets show the phase near
 $\bar{\vec{r}}=0$, with the winding number of (a) $1$, (b) $-1$, (c) $1$, and 
 (d) $0$. Parameters are set as $\Delta_0/V_C=0.25$ and 
 $n_x\times n_y=4\times8$.
 }
 \label{fig:PCF}
\end{figure}
To further support topological nature, we compute the pair amplitude,
\begin{align}
 F_{\sigma_1\sigma_2}(\vec{R},\bar{\vec{r}})=\left\langle
 c^\dagger_{\sigma_1}(\vec{R}+\bar{\vec{r}}/2)
 c^\dagger_{\sigma_2}(\vec{R}-\bar{\vec{r}}/2)
 \right\rangle
 \label{eq:PCF}
\end{align}
where $\vec{R}$ and $\bar{\vec{r}}$ are the center-of-mass and relative 
coordinates. Although 
the attached superconductor has $s$-wave pairing, 
spin-orbit coupling induces a mixture of $s$- and $p$-wave pairings.
Figure~\ref{fig:PCF} shows 
$F_{\sigma_1\sigma_2}(\vec{R},\bar{\vec{r}})$ decomposed into $p$-wave 
(total spin $s_z=-1,0,1$) and $s$-wave ($s_z=0$) channels. 
The phase of the $p$-wave pair amplitude shows a winding around the 
origin with
winding numbers $+1$ or $-1$ (see insets).
Since the real-space winding directly maps to
the momentum-space one (see \ref{sec:winding} in SM~\cite{SM}), these results 
demonstrate the
emergence of $p\pm ip$ structures~\cite{Read00,Mishmash19} in the RID-TSC 
phase.

\paragraph{Concluding remarks.}--- 
In this Letter, we studied a half-filled Rashba-coupled Landau level
proximity-coupled to an $s$-wave superconductor,
and demonstrated the emergence of the repulsive-interaction-driven topological
superconductivity.
Our work lends itself to generalization in many directions.
While our study focused on the filling factor $\nu=1/2$, extensions to other 
fillings, such as a hybrid with $\nu=2/3$ FQH state~\cite{Mong14},
present a natural direction for future work.
Coupling the edges of fractional quantum Hall states to superconductors 
has been proposed to yield various kinds of non-Abelian anyons~\cite{Cheng12,Lindner12,Burrello13,Clarke13,Vaezi13,Milsted14,Klinovaja14,Vaezi14,Mong14,Alicea16,Sagi17,Hu18,Repellin18,Lopes19,Liang19,Nielsen22,Schiller23}, suggesting that our 
bulk hybrid setting may likewise host rich and unexplored physics. 

By regarding the proximity effect as a mean-field treatment 
of intrinsic attractive interactions, our model reduces to a
problem of electrons with both repulsive and attractive interactions partially 
filling a topological flat band.
Very recently, chiral superconductivity has been observed in multilayer 
rhombohedral graphene~\cite{Han25,Morissette25}, whereas fractional quantum 
anomalous Hall effects appear with a moir\'e potential~\cite{Lu24}. This 
discovery has stimulated intensive 
theoretical work~\cite{Shi25,Divic25,Shi25b,Kim25,Shi25c,Nosov25,Pichler25,Wang25aa,Han25b,Wang25bb,Zhang25} to explore the interplay 
between quantum Hall physics and superconductivity.
It would also be intriguing to apply our model to this problem.

\begin{acknowledgments}
 K.K. thanks Songyang Pu, Naokazu Shibata, and Sora Araki for helpful 
 discussions.
 We acknowledge the computational resources offered by Research Institute for
 Information Technology, Kyushu University, 
 and the Supercomputer Center, the
 Institute for Solid State Physics, the University of Tokyo. 
 The work is supported in part by JSPS KAKENHI Grant nos. 
 JP23K19036, 
 JP24K06926, 
 JP25K17318, 
 JP25H01250, 
 JP25H00613. 
 J.K.J. acknowledge support in part by the U.S. National Science Foundation under Grant No. DMR-2404619.
\end{acknowledgments}

\bibliography{biblio_fqhe,biblio_fqhe_original,biblio_fqhe_1}

\begin{thebibliography}{125}
\bibitem [{scr()}]{screen}%
  \BibitemOpen
  \href@noop {} {}\bibinfo {note} {We choose a screened Coulomb interaction for
  two reasons. (i) As mentioned in the Introduction, a weakening or screening
  of the e-e interaction would be required to realize TSC, for example nearby
  screening layers~\cite{Nakamura20}. (ii) It avoids complications arising from
  the bare Coulomb potential. For $V(r)=1/r$, a positive background ensuring
  charge neutrality is normally required to remove the singular behavior of the
  Fourier component at $\vec{k}=0$. However, this treatment fails when working
  in the full Fock space: if $\vec{k}=0$ component is simply removed, states
  with larger electron number spuriously acquire lower energy due to the
  background contribution. This issue is absent for a screened interaction,
  which does not require introducing a positive background. For simplicity, we
  set the screening length to be the magnetic length.}\BibitemShut {Stop}%
\bibitem [{Abr()}]{Abrikosov}%
  \BibitemOpen
  \href@noop {} {}\bibinfo {note} {$\varphi^\tx{LLL}_{j}(\vec{r})$ is the
  unnormalized wavefunction of the lowest LL in the Landau gauge:
  $\varphi^\tx{LLL}_{j}(\vec{r})=\sum_{s=-\infty}^{\infty}
  e^{-i\left(k_j+sL_x/\bar{l}_\tx{B}^2\right)y
  -\left(k_j\bar{l}_\tx{B}^2+sL_x-x\right)^2/2\bar{l}_\tx{B}^2}$, where
  $k_j=2\pi j/a$ with $a$ the intervortex separation, $\bar{l}_{B}$ is the
  magnetic length for charge-$2e$ Cooper pairs, and the summation comes from
  the torus geometry of our system. Although the definition of $k_j$ can in
  principle be shifted by $2\pi/L_y$, we adopt the above convention so that the
  nodes of $\Delta_{\bm{k}}$ are included in the allowed momenta for finite
  systems.}\BibitemShut {Stop}%
\bibitem [{flu()}]{flux}%
  \BibitemOpen
  \href@noop {} {}\bibinfo {note} {The flux quantum for electrons is
  $\phi_0\equiv h/e$ while that for charge-$2e$ Cooper pairs is
  $\bar{\phi}_0\equiv h/2e=\phi_0/2$. Therefore, the system with $N_\phi$
  magnetic unit cells (i.e. the total magnetic flux is $N_\phi\phi_0$) contains
  $2N_\phi$ vortices.}\BibitemShut {Stop}%
\bibitem [{gR()}]{gR}%
  \BibitemOpen
  \href@noop {} {}\bibinfo {note} {The pairing matrix $\Delta_{\vec{k}}$ in
  Eq.~\eqref{eq:HDelta} becomes independent of the Rashba-coupling strength
  $g_R$ by rescaling
  $\Delta_0\rightarrow\Delta_0/\beta(g_R)$~\cite{Mishmash19}, where
  $\beta(g_R)$ is a monotonically increasing function with $\beta(0)=0$ and
  $\beta(\infty)=1/2$. We hence absorb this scaling into $\Delta_0$. When
  evaluating the interaction matrix elements
  $V_{\vec{k}_1\vec{k}_2\vec{k}'_1\vec{k}'_2}$ in Eq.~\eqref{eq:Hint}, we take
  the $g_R\rightarrow\infty$ limit for simplicity.}\BibitemShut {Stop}%
\bibitem [{Lan()}]{Lanczos}%
  \BibitemOpen
  \href@noop {} {}\bibinfo {note} {When computing the lowest $N$ eigenvalues
  using the Lanczos algorithm, the iteration is terminated once all targeted
  eigenvalues satisfy $|E(i)-E(i-3)|<10^{-8}$, where $E(i)$ denotes the
  eigenvalue obtained at the $i$th step. Gram-Schmidt orthonormalization is
  used at each iteration step to maintain orthogonality.}\BibitemShut {Stop}%
\bibitem [{Ras()}]{RashbaCF}%
  \BibitemOpen
  \href@noop {} {}\bibinfo {note} {Our target LL is identical to the $n=1$ LL
  of graphene with full spin and valley polarization. According to
  Refs.~\onlinecite{Toke07b,Wojs11a,Balram15c}, the fully spin- and
  valley-polarized composite Fermi liquid is favored over the Pfaffian state in
  the $n=1$ LL of graphene for the Coulomb interaction. Although we consider
  the screened Coulomb interaction, the qualitative conclusion is not expected
  to change.}\BibitemShut {Stop}%
\bibitem [{Nex()}]{NextSize}%
  \BibitemOpen
  \href@noop {} {}\bibinfo {note} {The largest accessible size,
  $(n_x,n_y)=(4,8)$, has the Hilbert space dimension of $6.7\times10^{7}$
  within a given $(\bm{K},P)$ sector (the dimensions are comparable across
  sectors). The next larger size, $(n_x,n_y)=(10,4)$, has the dimension of
  $1.4\times10^{10}$, which exceeds the practical limits of exact
  diagonalization.}\BibitemShut {Stop}%
\end{thebibliography}

\clearpage
\renewcommand{\thesection}{S\arabic{section}}
\renewcommand{\theequation}{S\arabic{equation}}
\renewcommand{\thefigure}{S\arabic{figure}}
\renewcommand{\thetable}{S\arabic{table}}
\setcounter{equation}{0}
\setcounter{figure}{0}
\setcounter{table}{0}
\makeatletter
\c@secnumdepth = 2
\onecolumngrid
\begin{center}
 \Large{Supplemental Material}
\end{center}
\setcounter{page}{1}
\twocolumngrid

\section{Two-dimensional system in magnetic fields on a torus}
\label{sec:2d}
\subsection{Landau levels}
\label{sec:Landau}
Here, we present the explicit form of the single-particle wavefunctions of the 
Landau levels (LLs)
in the torus geometry. The Hamiltonian of a particle with charge $e<0$ in a
uniform magnetic field $\vec{B}=(0,0,B)$ is given by
\begin{align}
 H_\tx{kin}=\frac{\pi^2}{2m},
 \label{eq:Hkin}
\end{align}
where $\vec{\pi}=\vec{p}+|e|\vec{A}$ with the Landau gauge $\vec{A}=(0,Bx,0)$. 
We impose the following quasiperiodic boundary 
conditions,~\cite{Yoshioka83,Yoshioka84b,Kato93}
\begin{align}
 &\varphi(x+L_x,y)
 =e^{-i\frac{L_xy}{\lB^2}}\varphi(x,y),\\
 &\varphi(x,y+L_y)=\varphi(x,y),
\end{align}
where $L_{x(y)}$ is the system length and $\lB=\sqrt{\hbar/|e|B}$ is the
magnetic length. The eigenfunction is given by
\begin{align}
 \varphi_{nj}(\vec{r})
 =&\mathcal{N}_n
 \sum_{s=-\infty}^{\infty}
 H_n\left(\frac{x-sL_x-k_j\lB^2}{\lB}\right)\times\non
 &\qquad\qquad
 e^{
 -i\left(
 k_j
 +\frac{sL_x}{\lB^2}\right)y
 -\frac{\left(k_j\lB^2+sL_x-x\right)^2}{2\lB^2}
 },
 \label{eq:varphi}
\end{align}
where $\mathcal{N}_n=\sqrt{1/L_yn!2^n\lB\sqrt{\pi}}$ and $H_n$ denotes the 
Hermite polynomials. Here, $n=0,1,2,\ldots$ labels the LL index, and 
$k_j=2\pi j/L_y$ with $j=0,1,\ldots,N_\phi-1$ is the wave 
number along the $y$-direction, where $N_\phi$ is the total number of magnetic
flux quanta threading the system

\subsection{Abrikosov vortex lattice}
\label{sec:Abri}
\begin{figure}[t]
\includegraphics[width=\columnwidth]{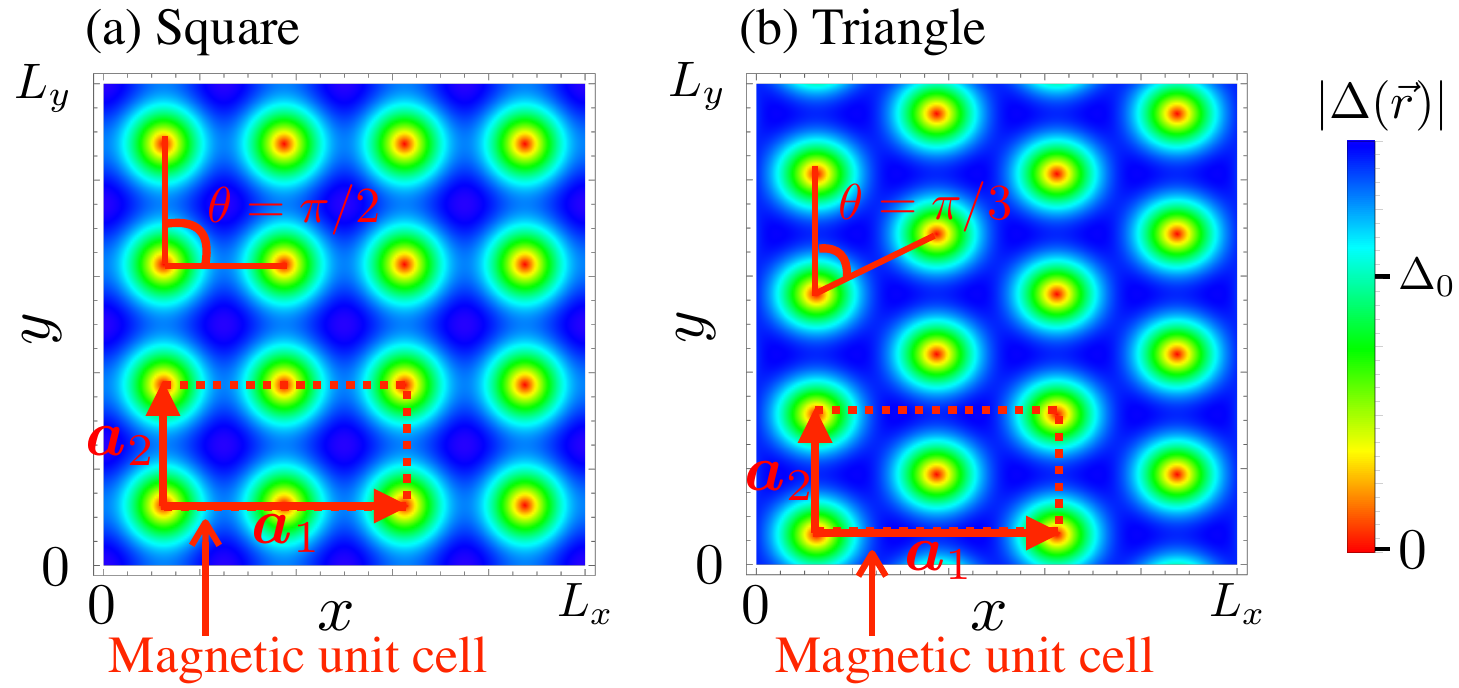}
 \caption{
 (a) Square and (b) triangular vortex lattices with the system size
 $(n_x,n_y)=(2,4)$. The 
 color indicates the modulus of the pair potential $\Delta(\vec{r})$, as
 defined in Eq.~\eqref{eq:Abri} with $C_0=\Delta_0/\sqrt{2}$. The dotted lines 
 indicate the magnetic unit cell. 
 }
 \label{fig:Abri}
\end{figure}

We derive the pair potential that describes vortex lattices. Here,
quantities associated with Cooper pairs are denoted with a bar, such as 
$\bar{l}_\tx{B}=\lB/\sqrt{2}$ and $\bar{N}_\phi=2N_\phi$. We denote by $\theta$
the angle between the primitive translation vectors of the vortex lattice;
$\theta=\pi/2$ $(\pi/3)$ for a square (triangular) lattice as shown in 
Fig.~\ref{fig:Abri}. The intervortex separation $a$ satisfies the relation
$a^2\s \bar{N}_{\phi}=L_xL_y$, which gives:
\begin{align}
 a=\lB\sqrt{\pi/\s}
\end{align}
where $\s$ is a shorthand of $\sin\theta$. 

Now we introduce a rectangular magnetic unit cell (MUC) as shown in 
Fig.~\ref{fig:Abri}. Let the number of MUCs be $n_x\times n_y$ with 
$n_x,n_y$ integers. The system lengths $L_x$ and $L_y$ satisfies 
$L_xL_y=2\pi\lB^2n_xn_y$ and $L_x/L_y=2\s n_x/n_y$. These yield
\begin{align}
 &L_x=n_x\sqrt{2\pi \lB^2\times2\s}=n_xQ\lB^2,\\
 &L_y=n_y\sqrt{2\pi \lB^2/\left(2\s\right)}=n_y\frac{2\pi}{Q},
\end{align}
where
\begin{align}
 Q=\frac{2\pi}{a}
\end{align}
We write the pair potential $\Delta(\vec{r})$ by summing over 
$\bar{\varphi}_{0j}$ with the wave numbers that are multiples of $Q$ 
as~\cite{Tinkham96,Mishmash19}
\begin{align}
 \Delta(\vec{r})
 &=\sum_{p=0}^{2n_x-1}C_{p}\sum_{s=-\infty}^{\infty}
 e^{
 -i\left(
 k_{pn_y}
 +\frac{sL_x}{\bar{l}_\tx{B}^2}\right)y
 -\frac{\left(k_{pn_y}\bar{l}_\tx{B}^2+sL_x-x\right)^2}
 {2\bar{l}_\tx{B}^2}
 }\non
 &=\sum_{p=0}^{2n_x-1}C_{p}\sum_{s=-\infty}^{\infty}
 e^{
 -i\left(
 k_{pn_y}
 +\frac{2sL_x}{\lB^2}\right)y
 -\frac{2\left(\frac{k_{pn_y}\lB^2}{2}+sL_x-x\right)^2}
 {2\lB^2}
 }.
 \label{eq:Abri}
\end{align}
The coefficients $C_p$ are given by
\begin{align}
 C_p=C_0e^{-i\pi p^2\cos\theta}
 =\left\{
 \begin{array}{ll}
  C_0 & \tx{(square)}\\
  C_0e^{-i\pi p^2/2} & \tx{(triangular)}
 \end{array}
 \right.
\end{align}
In the following, we set $C_0=\Delta_0/\sqrt{2}$, where $\Delta_0$ determines
the pairing strength. 

\subsection{Spin-orbit coupling}
\label{sec:soc}
Here we derives the spectrum of the Rashba-coupled 
LLs~\cite{Rashba60,Yu84,Schliemann03,Shen04,Ito12,Mishmash19,Chaudhary20}.
The following argument remains valid for other geometries, such as a
sphere. The Hamiltonian including Rashba spin-orbit coupling is given by
\begin{align}
  H'_\tx{kin}
  &=\frac{\pi^2}{2m}-\alpha_R(\vec{\sigma}\times\vec{\pi})_z\non
  &=\hbar\omega_c
  \left(
  \begin{array}{cc}
   a^\dagger a+\frac{1}{2} & -g_Ra \\
   -g_Ra^\dagger & a^\dagger a+\frac{1}{2}
  \end{array}
  \right).
\end{align}
where $\alpha_R$ is the Rashba coupling strength,
$g_R=\sqrt{2}\alpha_R/\lB\omega_c$, and 
$a^\dagger=-i(\pi_x+i\pi_y)\lB/\sqrt{2}\hbar$. We consider the subspace spanned
by the following basis for $n\geq1$:
\begin{align}
 \Psi_{nj}\equiv
 \left(
 \begin{array}{cc}
  \left(
   \begin{array}{c}
    \ket{n-1,j} \\
    0
   \end{array}
  \right) & 
  \left(
   \begin{array}{c}
    0 \\
    \ket{nj}
   \end{array}
  \right)
 \end{array}
 \right).
\end{align}
Here, $\ket{nj}$ denotes the state of the $n$th LL without spin-orbit coupling
characterized by a quantum number $j$.
Within this subspace, the Hamiltonian reduces to
 \begin{align}
  \Psi_{nj}^\dagger H'_\tx{kin}\Psi_{nj}
  &=\left(
  \begin{array}{cc}
   n-\frac{1}{2} & -g_R\sqrt{n} \\
   -g_R\sqrt{n} & n+\frac{1}{2}
  \end{array}
  \right).
 \end{align}
Its eigenvalues and eigenvectors are given by
\begin{align}
 \epsilon_{n,\tau}
 &=\hbar\omega_c  \left(n\tau\sqrt{\frac{1}{4}+g_R^2n}\right),\\
 \vec{v}_{n,\tau}
 &=\left(
 \begin{array}{c}
  \cos\left(\frac{\pi}{4}\alpha_{n,\tau}\right) \\
  \sin\left(\frac{\pi}{4}\alpha_{n,\tau}\right)
 \end{array}
 \right),
\end{align}
where $\tau=\pm1$ labels the two Rashba-split branches, and
\begin{align}
 \alpha_{n,\tau}
 =\frac{4}{\pi}\arctan
 \left\{\frac{-\frac12-\tau\sqrt{g_R^2n+\frac14}}{g_R\sqrt{n}}\right\}.
\end{align}
In addition, the unpaired state $\left(0,\ket{0j}\right)^T$ is also an 
eigenstate of $H'_\tx{kin}$ with energy $\epsilon_0=\hbar\omega_c/2$.

\subsection{Bloch basis}
\label{sec:Bloch}
Here we review the Bloch basis~\cite{Mishmash19} for the LLs.
While we assume no spin-orbit coupling, the argument can be extended to include
it. Using the pseudomomentum 
$\vec{\mathcal{K}}=\vec{p}+|e|\vec{A}-|e|\vec{B}\times\vec{r}$,
we define the magnetic translation operators:
\begin{align}
 &T_1\equiv
 e^{-\frac{i}{\hbar}\vec{a}_1\cdot\vec{\mathcal{K}}},\\
 &T_2\equiv
 e^{-\frac{i}{\hbar}\vec{a}_2\cdot\vec{\mathcal{K}}},
\end{align}
where $\vec{a}_1=2a\s(1,0,0)=Q\lB^2(1,0,0)$ and $\vec{a}_2=a(0,1,0)$ are the 
primitive 
vectors of the MUC as shown in Fig.~\ref{fig:Abri}. Although these operators
commutes with $H_\tx{kin}$ in Eq.~\eqref{eq:Hkin}, the single-particle 
wavefunction $\varphi_{nj}$ in
Eq.~\eqref{eq:varphi} is not an eigenstate of $T_1$:
\begin{align}
 &T_1\ket{\varphi_{nj}}=\ket{\varphi_{n,j+n_y}},\\
 &T_2\ket{\varphi_{nj}}=e^{ik_ja}\ket{\varphi_{nj}}.
\end{align}
To construct a basis that diagonalizes both $T_1$ and $T_2$ simultaneously, we
define
\begin{align}
 \ket{\phi_{n\vec{j}}}
 &=\frac{1}{\sqrt{n_x}}\sum_{r=0}^{n_x-1}
 \left(
 e^{-ik_{j_x}Q\lB^2}T_1\right)^r
 \ket{\varphi_{nj_y}}\non
 &=\frac{1}{\sqrt{n_x}}\sum_{r=0}^{n_x-1}
 e^{-ik_{j_x}Q\lB^2r}
 \ket{\varphi_{n,j_y+rn_y}},
 \label{eq:phi}
\end{align}
where $k_{j_x}=2\pi j_x/L_x$. Here, $\vec{j}=(j_x,j_y)$ labels the Bloch
momentum in the magnetic Brillouin zone, where 
$j_{x(y)}=0,1,\ldots,n_{x(y)}-1$. By definition, this Bloch basis satisfies
\begin{align}
 T_1\ket{\phi_{n\vec{j}}}
 &=e^{ik_{j_x}Q\lB^2}\ket{\phi_{n\vec{j}}},\\
 T_2\ket{\phi_{n\vec{j}}}
 &=e^{ik_{j_y}a}\ket{\phi_{n\vec{j}}}.
\end{align}

\section{Matrix elements}
\label{sec:matrix}
In this section, we derive the projected Hamiltonian in Eqs.~\eqref{eq:Hint} 
and \eqref{eq:HDelta} using the Bloch basis.
We begin by defining the real-space integration, which appears repeatedly
in the following discussion:
\begin{align}
 &\int d^2r\,f(\vec{r})\non
 =&\sum_{s_x=-\infty}^{\infty}\sum_{s_y=-\infty}^{\infty}
 \int_0^{L_x} dx\int_0^{L_y} dyf(x+s_xL_x,y+s_yL_y),
\end{align}
where $L_{x(y)}$ is the system lengths in the $x(y)$-direction.

\subsection{Interaction}
Let us recall the interaction part of the Hamiltonian in 
Eq.~\eqref{eq:originalH}:
\begin{align*}
 &H_\tx{int}
 =\frac{1}{2}\sum_{\sigma_1\sigma_2}\int d^2r_1d^2r_2\,
 c_{\sigma_1}^\dagger(\vec{r}_1)c_{\sigma_2}^\dagger(\vec{r}_2)
 V(r_1-r_2)
 c_{\sigma_2}(\vec{r}_2)c_{\sigma_1}(\vec{r}_1).
\end{align*}
We first consider the case without spin-orbit coupling. 
Using the Bloch basis 
defined in Eq.~\eqref{eq:phi}, the Hamiltonian becomes
\begin{align}
 H_\tx{int}
 =&\frac{1}{2}
 \sum_{\sigma_1\sigma_2}
 \sum_{n_1n_2n_1'n_2'}
 \sum_{\vec{j}_1\vec{j}_2\vec{j}'_{1}\vec{j}'_{2}}
 V_{\vec{j}_1\vec{j}_2;\vec{j}'_{1}\vec{j}'_{2}}^{(n_1n_2n_1'n_2')}\times\non
 &\qquad\qquad\qquad
 c_{\sigma_1n_1\vec{j}_1}^\dagger c_{\sigma_2n_2\vec{j}_2}^\dagger
 c_{\sigma_2n_2'\vec{j}'_2}c_{\sigma_1n_1'\vec{j}'_1}\non
 \rightarrow
 &\frac{1}{2}
 \sum_{\sigma_1\sigma_2}
 \sum_{n_1n_2}
 \sum_{\vec{j}_1\vec{j}_2\vec{j}'_{1}\vec{j}'_{2}}
 V_{\vec{j}_1\vec{j}_2;\vec{j}'_{1}\vec{j}'_{2}}^{(n_1n_2)}\times\non
 &\qquad\qquad\qquad
 c_{\sigma_1n_1\vec{j}_1}^\dagger c_{\sigma_2n_2\vec{j}_2}^\dagger
 c_{\sigma_2n_2\vec{j}'_2}c_{\sigma_1n_1\vec{j}'_1},
\end{align}
where $c_{n\vec{j}}^\dagger$ creates a fermion in the Bloch state
$\phi_{n\vec{j}}$. In the second line, in preparation for the projection to a 
Rashba-coupled LL
(performed below), we omit terms that are killed by that projection. The matrix element is given by
\begin{widetext}
\begin{align}
 V_{\vec{j}_1\vec{j}_2;\vec{j}_{1'}\vec{j}_{2'}}^{(n_1n_2)}
 &\equiv\left(\bra{\phi_{n_1\vec{j}_1}}\otimes\bra{\phi_{n_2\vec{j}_2}}\right)
 V	
 \left(\ket{\phi_{n_1\vec{j}'_{1}}}\otimes\ket{\phi_{n_2\vec{j}'_{2}}}\right)
 \non
 &=
 \delta^{\tx{mod}}
 _{\vec{j}_{1}+\vec{j}_{2}-\vec{j}'_{1}-\vec{j}'_{2},\vec{0}}
 \frac{1}{L_xL_y}
 \frac{1}{n_x}
 e^{-i
 k_{j'_{1x}}\left(k_{j_{1y}}+k_{j_{2y}}-k_{j'_{1y}}-k_{j'_{2y}}\right)
 \lB^2}
 \sum_{s,t=0}^{n_x-1}
 e^{i\left(
 \left(k_{j_{1x}}-k_{j'_{1x}}\right)s
 +\left(k_{j_{x2}}-k_{j'_{1x}}\right)t
 \right)Q\lB^2}\non
 &\qquad
 \times
 \sum_{i_x,i_y=-\infty}^{\infty}V\left(\vec{q}_{\vec{i}}\right)
 \delta^{\tx{mod} N_\phi}_{j_{2y}-j'_{2y}+tn_y-i_y,0}
 e^{i\left(k_{j_{1y}}-k_{j'_{2y}}\right)q_{i_x}\lB^2
 +iQ\lB^2sq_{i_x}
 -\frac{q^2\lB^2}{2}}
 L_{n_1}\left(\frac{q^2\lB^2}{2}\right)
 L_{n_2}\left(\frac{q^2\lB^2}{2}\right),
 \label{eq:Vjjjj}
\end{align}
\end{widetext}
where $L_n$ are the Laguerre polynomials, $V(\vec{q})$ is the Fourier 
transformation of the interaction $V(\vec{r})$,
$\vec{q}_{\vec{i}}=(q_{i_x},q_{i_y})$ with 
$q_{i_\alpha}=2\pi i_\alpha/L_\alpha$, and
$\delta^\tx{mod}_{\vec{j}\vec{j}'}=\delta^{\tx{mod}\,n_x}_{j_xj_x'}\delta^{\tx{mod}\,n_y}_{j_yj_y'}$. 
Here, $\delta^{\tx{mod}\,n}_{jj'}=1$ if
$j\equiv j'(\tx{mod}\,n)$, and $0$ otherwise.
The matrix elements using the Landau basis are given in 
Ref.~\onlinecite{Yoshioka83}. Our result in Eq.~\eqref{eq:Vjjjj} provides their
counterpart in the Bloch basis.

Now, we consider the case with spin-orbit coupling. Projecting the interaction
onto the Rashba-coupled LL with energy $\epsilon_{n,\tau}$, the Hamiltonian
becomes
\begin{align}
 \tilde{H}_\tx{int}
 &=\frac{1}{2}
 \sum_{\vec{j}_1\vec{j}_2\vec{j}'_1\vec{j}'_2}
 V^{(n\tau)}_{\vec{j}_1\vec{j}_2;\vec{j}'_1\vec{j}'_2}
 c^\dagger_{n\tau\vec{j}_1}c^\dagger_{n\tau\vec{j}_2}
 c_{n\tau\vec{j}'_2}c_{n\tau\vec{j}'_1},
\end{align}
where 
\begin{widetext}
\begin{align}
 V^{(n\tau)}_{\vec{j}_1\vec{j}_2;\vec{j}'_1\vec{j}'_2}
 &=
 \delta^{\tx{mod}}
 _{\vec{j}_{1}+\vec{j}_{2}-\vec{j}'_{1}-\vec{j}'_{2},\vec{0}}
 \frac{1}{L_xL_y}
 \frac{1}{n_x}
 e^{-i
 k_{j'_{1x}}\left(k_{j_{1y}}+k_{j_{2y}}-k_{j'_{1y}}-k_{j'_{2y}}\right)
 \lB^2}
 \sum_{s,t=0}^{n_x-1}
 e^{i\left(
 \left(k_{j_{1x}}-k_{j'_{1x}}\right)s
 +\left(k_{j_{x2}}-k_{j'_{1x}}\right)t
 \right)Q\lB^2}\non
 &\quad
 \times
 \sum_{i_x,i_y=-\infty}^{\infty}V\left(\vec{q}_{\vec{i}}\right)
 \delta^{\tx{mod} N_\phi}_{j_{2y}-j'_{2y}+tn_y-i_y,0}
 e^{i\left(k_{j_{1y}}-k_{j'_{2y}}\right)q_{i_x}\lB^2
 +iQ\lB^2sq_{i_x}
 -\frac{q^2\lB^2}{2}}
 \left(F_{n\tau}\left(q\right)\right)^2.
\end{align}
\end{widetext}
Here,
\begin{align}
 F_{n\tau}(q)&=
 \left|\left[\vec{v}_{n\tau}\right]_{\uparrow}\right|^2
 L_{n-1}\left(\frac{q^2\lB^2}{2}\right)\non
 &\qquad\qquad
 +\left|\left[\vec{v}_{n\tau}\right]_{\downarrow}\right|^2
 L_{n}\left(\frac{q^2\lB^2}{2}\right),
\end{align}
and $\left[\vec{v}_{n\tau}\right]_{\uparrow(\downarrow)}
=\left[\vec{v}_{n\tau}\right]_{1(2)}$.

\subsection{Pair potential}
We now turn to the pairing term in the Hamiltonian in 
Eq.~\eqref{eq:originalH}:
\begin{align*}
 &H_\Delta
 =\int d^2r
 \left[
 c^\dagger_\uparrow(\vec{r})\Delta(\vec{r})c^\dagger_\downarrow(\vec{r})
 +\tx{h.c.}\right],
\end{align*}
where $\Delta(\vec{r})$ is given by Eq.~\eqref{eq:Abri}. 
We first consider the case without spin-orbit coupling. Using the Bloch basis 
defined in Eq.~\eqref{eq:phi}, the pairing Hamiltonian can be written as
\begin{align}
 H_\Delta
 &= \sum_{n_1n_2}
 \sum_{\vec{j}_1\vec{j}_2}
 \Delta_{n_1\vec{j}_1;n_2\vec{j}_2}
 c_{\uparrow n_1\vec{j}_1}^\dagger c_{\downarrow n_2\vec{j}_2}^\dagger
 +\tx{h.c.}
\end{align}
Here, the matrix element is given by
\begin{widetext}
\begin{align}
 &\Delta_{n_1\vec{j}_1;n_2\vec{j}_2}
 \equiv
 \int\,d^2r\,\phi_{n_2\vec{j}_2}^*(\vec{r})
 \phi_{n_1\vec{j}_1}^*(\vec{r})\Delta(\vec{r})\non
 &
 =\delta^{\tx{mod}}_{\vec{j}_1+\vec{j}_2,0}
 \Delta_0A_{n_1n_2}\sum_{u=0}^{2n_x-1}e^{-i\pi u^2\cos\theta
 -ik_{j_{x2}}(uQ+k_{j_{y1}}+k_{j_{y2}})\lB^2}
 \sum_{s=-\infty}^{\infty}\delta^{\tx{mod}2}_{s,0}
 e^{-\frac{\left(2k_{j_{y1}}\lB^2
 +uQ\lB^2
 -sL_x\right)^2}{4\lB^2}
 }
 H_{n_1+n_2}\left(
 \frac{2k_{j_{y1}}\lB^2
 +uQ\lB^2-sL_x
 }{\lB\sqrt{2}}\right)
 \end{align}
\end{widetext}
where $A_{n_1n_2}=\frac{(-1)^{n_1}}{2^{n_1+n_2}\sqrt{n_1!n_2!}}$, and
$H_n$ denotes the Hermite polynomial.
Here, the momentum index $j_{\alpha=x,y}$ takes values $0,1,\ldots,n_\alpha-1$.
For notational convenience, we denote by ``$-j_\alpha$'' the momentum index 
that satisfies $j_\alpha+(-j_\alpha)\equiv0\,\mod n_\alpha$.
Using this notation, the 
pairing Hamiltonian can be recast in the more compact form:
\begin{align}
 H_\Delta
 &= \sum_{n_1n_2}
 \sum_{\vec{j}}
 \Delta^{(n_1n_2)}_{\vec{j}}
 c_{\uparrow n_1\vec{j}}^\dagger c_{\downarrow n_2-\vec{j}}^\dagger+\tx{h.c.},
\end{align}
where $\Delta^{(n_1n_2)}_\vec{j}=\Delta_{n_1\vec{j};n_2-\vec{j}}$. In this
work, we compute $\Delta^{(n_1n_2)}_\vec{j}$ on the torus geometry. Although
similar results can also be obtained on the infinite cylinder 
geometry~\cite{Mishmash19}, the expressions are not exactly identical.

\begin{figure}[t]
\includegraphics[width=\columnwidth]{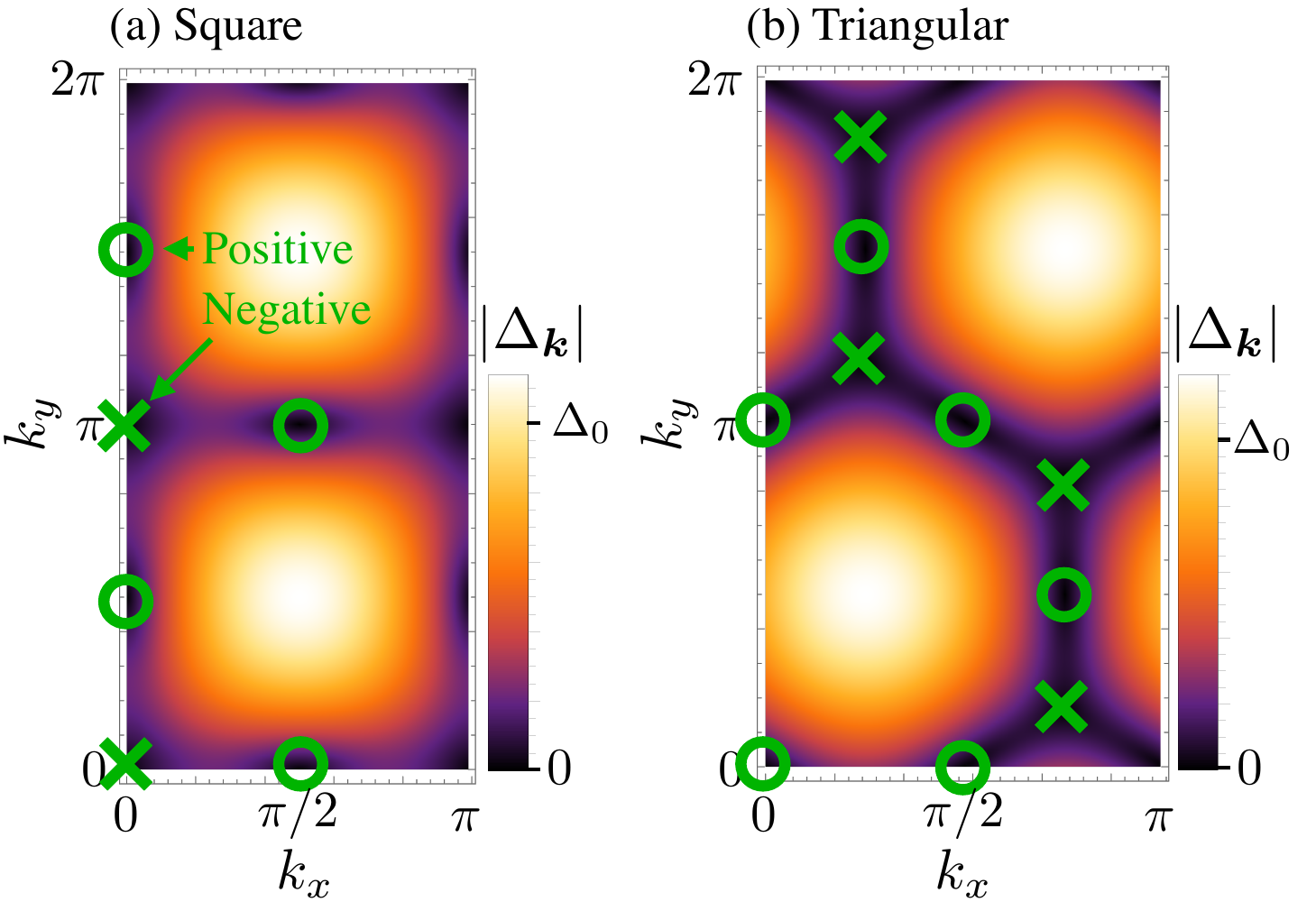}
 \caption{
 Modulus of $\Delta_{\vec{k}}\equiv\Delta^{(1-)}_{\vec{j}}$ 
 for (a) square and (b) triangular vortex lattices. Here, the wave number 
 $\vec{k}=(k_x,k_y)$ is defined using its index $\vec{j}$ as
 $k_\alpha=2\pi j_\alpha/L_\alpha$. Here, the intervortex separation $a$ is set
 to unity.
 The circles (crosses) indicate Dirac nodes with positive
 (negative) chirality.
 }
 \label{fig:nodes}
\end{figure}
Now, we introduce spin-orbit coupling. The Hamiltonian projected onto the
the Rashba-coupled LL with energy $\epsilon_{n,\tau}$ is given by
\begin{align}
 \tilde{H}_\Delta
 &=\sum_{\vec{j}}
 \Delta^{(n\tau)}_{\vec{j}}
 c_{n\tau\vec{j}}^\dagger c_{n\tau-\vec{j}}^\dagger+\tx{h.c.},
\end{align}
where the projected pairing matrix element is
\begin{align}
 \Delta^{(n\tau)}_{\vec{j}}
 =\left[\vec{v}_{n\tau}\right]^*_{\uparrow}
 \left[\vec{v}_{n\tau}\right]^*_{\downarrow}\Delta^{(n-1,n)}_\vec{j}.
\end{align}
In Fig.~\ref{fig:nodes}, we plot the modulus $|\Delta^{(1-)}_{\vec{j}}|$ as a 
function of the momentum $\vec{k}=(k_x,k_y)=(2\pi j_x/L_x,2\pi j_y/L_y)$. Each 
Dirac node is characterized by a positive or negative 
chirality~\cite{Mishmash19}. 
For the square vortex lattice, there are six Dirac nodes. As mentioned 
in the main text, the system size must be of the form
$n_x\times n_y=2s\times4t$ (with $s,t$ integers) to ensure that all nodes are 
included within the discrete Brillouin zone.

\begin{figure*}[t]
\includegraphics[width=2\columnwidth]{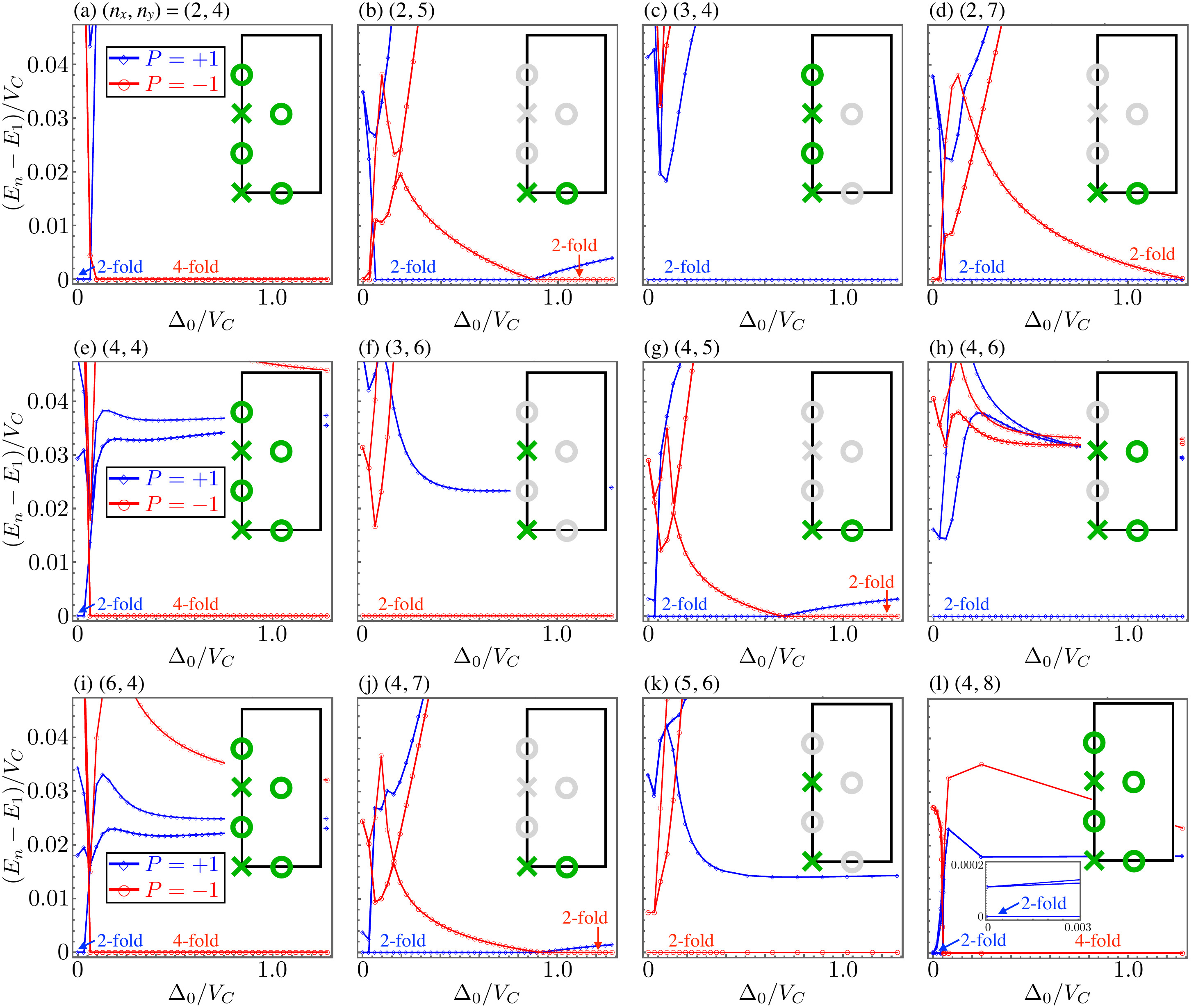}
 \caption{
 Same as Fig.~\ref{fig:CFStoSC}(a) in the main text, but for various system
 sizes $(n_x,n_y)$ and a wider $\Delta_0/V_C$ range. Here, $\Delta_0$ is the 
 pairing strength, $V_C$ is the interaction strength, and $E_i$ is the $i$th 
 lowest energy. We plot the lowest five energy states for each fermion 
 parity. The insets show the Dirac nodes of $\Delta_\vec{k}$ in
 Fig.~\ref{fig:nodes}(a), where green (gray) markers indicate that the
 corresponding node is (is not) included in the discrete Brillouin zone. 
 Each panel corresponds to a system size $(n_x,n_y)$ satisfying either 
 $(n_x,n_y)=(2s,4t)$ with $s,t$ integers or $0.5\leq L_x/L_y\leq2$, where 
 $L_{x(y)}$ is the system length in $x(y)$-direction. In (l), fewer data 
 points are shown since this calculation requires more computational resources.
 The inset in (l) is a zoomed-in plot.
 }
 \label{fig:nxny}
\end{figure*}
\clearpage
\section{Energy spectra for various $(n_x,n_y)$}
\label{sec:nxny}
Here we discuss the system-size dependence of the many-body energy spectrum. 
Figure~\ref{fig:nxny} shows energy spectra for various system sizes 
as functions of the pairing strength, analogous to Fig.~\ref{fig:CFStoSC}(a) in
the main text.

We find that the spectral features strongly depend on which Dirac nodes of 
$\Delta_\vec{k}$ are included in the discrete Brillouin zone. The observed
behavior can be categorized into the following three cases:
\begin{enumerate}[(1)]
 \setlength{\parskip}{0cm}
 \setlength{\itemsep}{0cm}
 \item All nodes included:
       
       [Figs.~\ref{fig:nxny}(a)(e)(i)(l)]
       The system size is $(n_x,n_y)=(2s,4t)$ with $s,t$ integers.
       At small finite $\Delta_0/V_C$, the ground state is twofold degenerate 
       with even fermion parity, $P=+1$. As $\Delta_0/V_C$ increases, a 
       transition occurs 
       to a fourfold degenerate ground state with odd fermion parity, $P=-1$. 
       These
       states consistently appear at the total momentum $\vec{K}=(\pi/2,\pi/2)$
       and $(\pi/2,3\pi/2)$, with two state at each momentum.
       \label{item:allnode}

 \item Only two nodes at $\vec{k}=(0,0)$ and $(\pi/2,0)$ included:
       
       [Figs.~\ref{fig:nxny}(b)(d)(g)(j)]
       At small $\Delta_0/V_C$, the ground state is twofold degenerate 
       with
       even fermion parity, $P=+1$. As $\Delta_0$ increases, the fermion 
       parity of the ground state changes to odd, $P=-1$.
       \label{item:2nodes}

 \item Other cases
       
       [Figs.~\ref{fig:nxny}(c)(f)(h)(k)]
       The ground state remains twofold degenerate over the entire range of
       $\Delta_0/V_C$.
       \label{item:nonode}
\end{enumerate}

This dependence of the ground state suggests that the low-energy physics is 
governed by the Dirac 
nodes of $\Delta(\vec{k})$. In the thermodynamic limit, the Brillouin zone
becomes continuous and includes all Dirac nodes. Therefore, case 
(\ref{item:allnode}) provides a finite-size analog that faithfully reflects
the the system in the thermodynamic limit.
For this reason, we restrict the main text discussion to this case.

\section{Energy difference}
\label{sec:gap3-1}
\begin{figure}[t]
\includegraphics[width=\columnwidth]{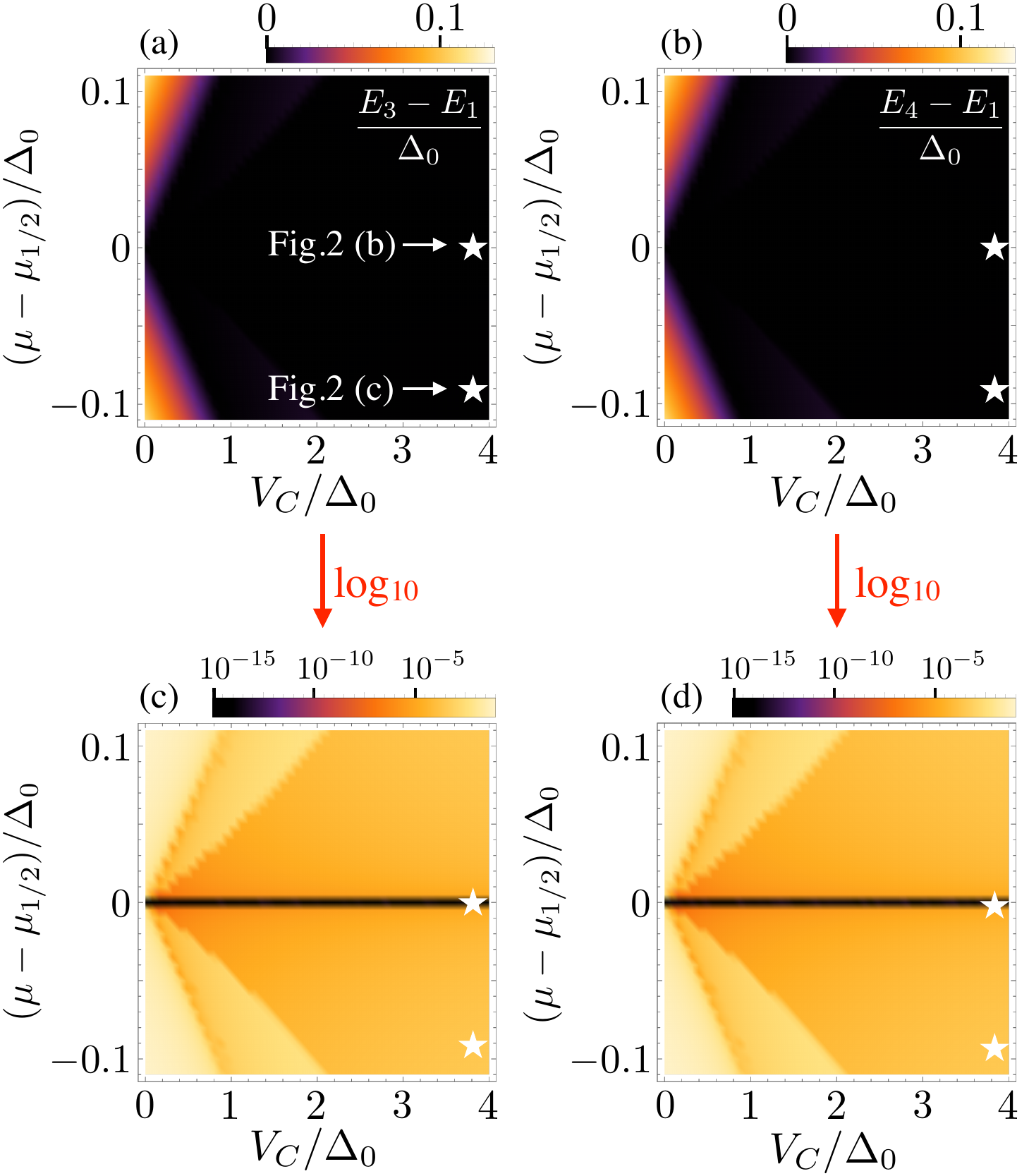}
 \caption{
 (a)(b) Same as Fig.~\ref{fig:Many-body} but for $(E_n-E_1)/\Delta_0$ with 
 $n=3,4$. (c)(d) Logarithmic scale plots of panels (a) and (b).
 }
 \label{fig:Many-bodySM}
\end{figure}
Figures~\ref{fig:Many-bodySM}(a)(b) are the same as Fig.~\ref{fig:Many-body} 
but for $(E_n-E_1)/\Delta_0$ with $n=3,4$. As mentioned in the main text, when 
$\nu$ deviates from $1/2$, the fourfold degeneracy of the ground state 
is lifted and only two degenerate states remain. Consequently, the splitting 
$E_3-E_1$ becomes nonzero. However, this energy splitting is too small to be 
resolved in Fig.~\ref{fig:Many-bodySM}(a). To visualize the behavior of the 
energy difference more clearly, we present logarithmic plots in 
Figs.~\ref{fig:Many-bodySM}(c)(d). 

\section{Fermion parity and the BdG Chern number}
\label{sec:FP}
Here we consider a generic BdG Hamiltonian for two-dimensional spinless 
fermions:
\begin{align*}
 H_\tx{BdG}=
 \sum_{\vec{k}}\epsilon_{\vec{k}}c^\dagger_\vec{k}c_\vec{k}-\mu N
 +\sum_\vec{k}\Delta_\vec{k}c^\dagger_\vec{k}c^\dagger_{-\vec{k}}+\tx{h.c.},
\end{align*}
where $\epsilon_\vec{k}$ is the single-particle dispersion (assumed symmetric,
$\epsilon_\vec{k}=\epsilon_{-\vec{k}}$) and
$\Delta_{\vec{k}}$, satisfying $\Delta_{-\vec{k}}=-\Delta_{\vec{k}}$, is the 
pair potential. 
Since $H_\tx{BdG}$ contains no 
interactions, it reduces to the standard BdG form, up to an irrelevant 
constant:
\begin{align}
 H_\tx{BdG}
 =\frac{1}{2}\sum_{\vec{k}}
 \left(c^\dagger_\vec{k}, c_\vec{-k}\right)
 h_\tx{BdG}(\vec{k})
 \left(
 \begin{array}{c}
  c_\vec{k} \\
  c^\dagger_{-\vec{k}}
 \end{array}
 \right),
 \label{eq:Href}
\end{align}
with
\begin{align}
 h_\tx{BdG}(\vec{k})
 =\left(
 \begin{array}{cc}
  \xi_\vec{k} & 2\Delta_\vec{k} \\
  2\Delta^*_\vec{k} & -\xi_{-\vec{k}}
 \end{array}
 \right),\ 
 \xi_\vec{k}=\epsilon_{\vec{k}}-\mu
\end{align}
Particle-hole symmetry implies
\begin{align}
 \Xi^{-1}h_\tx{BdG}(-\vec{k})\Xi
 =-h_\tx{BdG}(\vec{k}),
 \label{eq:PHsym}
\end{align}
where 
\begin{align}
 \Xi=\left(
 \begin{array}{cc}
  0 & 1 \\
  1 & 0
 \end{array}
 \right)K,
\end{align}
and $K$ is complex conjugation. We introduce quasiparticle operators:
\begin{align}
 \alpha_\vec{k}^\dagger
 =\left(
 c_\vec{k}^\dagger,c_{-\vec{k}}\right)\left(
 \begin{array}{c}
  u_\vec{k} \\
  v_\vec{k}
 \end{array}
 \right)
 =u_\vec{k}c^\dagger_\vec{k}+v_\vec{k}c_{-\vec{k}},
\end{align}
where $u_\vec{k}$ taken real without loss of generality.
The condition $[H_\tx{BdG},\alpha_\vec{k}]=0$ reduces to the eigenvalue
equation
\begin{align}
 h_\tx{BdG}(\vec{k})
 \left(
 \begin{array}{c}
  u_\vec{k} \\
  v_\vec{k}
 \end{array}
 \right)
 =E_\vec{k} \left(
 \begin{array}{c}
  u_\vec{k} \\
  v_\vec{k}
 \end{array}
 \right).
\end{align}
The energy spectrum is
\begin{align}
 E^\pm_\vec{k}
 =\pm\sqrt{\xi_{\vec{k}}^{2}+|2\Delta_\vec{k}|^2},
\end{align}
The eigenvector for $E^+_\vec{k}$ satisfies
\begin{align}
 |u_\vec{k}|^2
 &=\frac{1}{2}\left(1+\frac{\xi_{\vec{k}}}
 {\sqrt{\xi_{\vec{k}}^{2}+|2\Delta_\vec{k}|}}\right),\non
 |v_\vec{k}|^2
 &=\frac{1}{2}\left(1-\frac{\xi_{\vec{k}}}
 {\sqrt{\xi_{\vec{k}}^{2}+|2\Delta_\vec{k}|}}\right),\non
 \frac{v_\vec{k}}{u_\vec{k}}
 &=\frac{\sqrt{\xi_{\vec{k}}^{2}+|2\Delta_\vec{k}|}-\xi_{\vec{k}}}
 {2\Delta_\vec{k}^*}.
\end{align}
The particle-hole symmetry guarantees that the 
eigenvector of $E^-_\vec{k}$ given by $(v^*_{-\vec{k}},u^*_{-\vec{k}})^T$.
Since
\begin{align}
 \left(
 c_\vec{k}^\dagger,c_{-\vec{k}}\right)
 \left(
 \begin{array}{c}
  v^*_{-\vec{k}} \\
  u^*_{-\vec{k}}
 \end{array}
 \right)
 =\alpha_{-\vec{k}},
\end{align}
$H_\tx{BdG}$ can be written in terms of quasiparticles as
\begin{align}
 H_\tx{BdG}
 &=\frac{1}{2}\sum_\vec{k}\left(
 E_\vec{k}^+\alpha_\vec{k}^\dagger\alpha_\vec{k}
 +E_\vec{k}^-\alpha_\vec{-k}\alpha_\vec{-k}^\dagger
 \right)\non
 &=\frac{1}{2}\sum_\vec{k}\left(
 E_\vec{k}^+-E_\vec{-k}^-
 \right)\alpha_\vec{k}^\dagger\alpha_\vec{k}\non
 &=\sum_\vec{k}
 E_\vec{k}^+
 \alpha_\vec{k}^\dagger\alpha_\vec{k},
\end{align}
where we ignore an additive constant.

The ground state $\ket{\Omega}$ must satisfy, for all $\vec{k}$,
\begin{align}
 \alpha_\vec{k}\ket{\Omega}=0.
\end{align}
For $\vec{k}$ not an inversion-invariant momentum
(IIM, i.e. $\vec{k}=-\vec{k}$ modulo the Brillouin zone), the following
relation holds:
\begin{align}
 \alpha_{\pm\vec{k}}
 \left(u_\vec{k}^*-v_\vec{k}^*c_\vec{k}^\dagger c_{-\vec{k}}^\dagger\right)
 \ket{0}
 =&u_\vec{k}^*v_\vec{k}^*
 \left(-c_{\pm\vec{k}}c_\vec{k}^\dagger c_{-\vec{k}}^\dagger
 \pm c_{\mp\vec{k}}^\dagger\right)
 \ket{0}\non
 =&0,
\end{align}
where we used $u_{-\vec{k}}=u_{\vec{k}}$ and $v_\vec{-k}=-v_\vec{k}$. Here,
$\ket{0}$ denotes the electron vacuum. At nodes of $\Delta_{\vec{k}}$, denoted
$\vec{k}^*$ (including IIMs), the quasiparticle operator reduces to
\begin{align}
 \alpha_{\vec{k}^*}=\left\{
 \begin{array}{ll}
  c_{\vec{k}^*} & \tx{for $\xi_{\vec{k}^*}>0$},\\
  c_{-\vec{k}^*}^\dagger & \tx{for $\xi_{\vec{k}^*}<0$}.
 \end{array}
 \right.
\end{align}
Thus, the contribution of $\vec{k}^*$ to the ground state is
$\prod_{\xi_{\vec{k}^*}<0}c^\dagger_{-\vec{k}^*}\ket{0}$. This is equivalent to
$\prod_{\xi_{\vec{k}^*}<0}c^\dagger_{\vec{k}^*}\ket{0}$ since
$\xi_{\vec{k}}=\xi_{-\vec{k}}$. Combining these
results, the ground state takes the form
\begin{align}
 \ket{\Omega}
 =\prod_{\vec{k}\neq\vec{k}^*}{}'
 (u_\vec{k}+v_\vec{k}c^\dagger_\vec{k}c_{-\vec{k}}^\dagger)
 \prod_{\xi_{\vec{k}^*}<0}c^\dagger_{\vec{k}^*}\ket{0},
\end{align}
where 
$\prod{}'_{\vec{k}\neq\vec{k}^*}$ denotes the product over distinct
$(\vec{k},-\vec{k})$ pairs, excluding $\vec{k}^*$. This implies that the 
fermion parity of the ground state is determined solely by the 
occupations at $\vec{k}^*$'s.

Let us show that $P$ coincides with the parity of the BdG Chern number 
$\mathcal{N}$:
\begin{align}
 P=(-1)^\mathcal{N}.
 \label{eq:PNformulaSM}
\end{align}
In the limit $\mu\ll0$, where $|u_\vec{k}|^2\rightarrow1$ and 
$|v_\vec{k}|^2\rightarrow0$, the ground state reduces to the vacuum with 
$(P,\mathcal{N})=(+1,0)$, consistent with Eq.~\eqref{eq:PNformulaSM}. As $\mu$ 
increases, the gap closes whenever there exists a momentum $\vec{k}$ with 
$E^+_\vec{k}=0$. Since this condition requires $\Delta_\vec{k}=0$, the gap 
closing can occur only at nodes $\vec{k}^*$ where $\xi^+_{\vec{k}^*}=0$. 
Such a gap-closing changes the BdG Chern number by
$\mathcal{N}\rightarrow\mathcal{N}+n$, where $n$ is the winding number 
of $\Delta_\vec{k}$ expanded around the node,
and flips the fermion parity $P\rightarrow -P$.
Note that $n$ is an odd integer owing to $\Delta_{-\vec{k}}=-\Delta_{\vec{k}}$.
When gap closings occur at multiple points $\vec{k}^*$, the total change is the
sum of the individual contributions. The
resulting $(P,\mathcal{N})$ thus continues to satisfy 
Eq.~\eqref{eq:PNformulaSM}.

Since $\xi_{\vec{k}}=\xi_{-\vec{k}}$, any gap closing occurs simultaneously at 
$\vec{k}^*$ and $-\vec{k}^*$. Thus, only IIMs can host an individual gap closing, and consequently the total momentum of $\ket{\Omega}$ can take only IIM
values.

\section{Winding number}
\label{sec:winding}
The phase winding of a function $F(\vec{k})\propto k_x\pm ik_y$ around the 
origin can be inferred from the winding of its Fourier transform. This follows 
from
\begin{align}
 &\frac{1}{(2\pi)^2}\int dk^2\left(k_x\pm ik_y\right)
 e^{i\vec{k}\cdot\vec{r}}\non
 =&-i(\pa_x\pm i\pa_y)\delta^2(\vec{r})\non
 =&-i(\pa_x\pm i\pa_y)\lim_{\sigma\rightarrow0}
 \frac{1}{2\pi\sigma^2}\exp{-\frac{x^2+y^2}{2\sigma^2}}\non
 =&i(x\pm iy)\lim_{\sigma\rightarrow0}
 \frac{1}{2\pi\sigma^4}\exp{-\frac{x^2+y^2}{2\sigma^2}}.
\end{align}

\end{document}